\documentclass[useAMS,usenatbib]{mn2e}
\usepackage{txfonts}
\usepackage{float}
\usepackage{amssymb}
\usepackage{epsfig}
\usepackage{color}
\usepackage{graphicx}
\usepackage{multirow}
\usepackage{longtable}
\usepackage{verbatim}
\usepackage{caption}
\usepackage{array}

\title[Cross-correlations depending on luminosity in XTE~J1701-462]
{
\begin{center}
Cross-correlations
between soft and hard light curves depending
on luminosity in the transient neutron star XTE~J1701-462
\end{center}
}
\author[Y. N. Wang et al.]{Y. N. Wang$^{123}$\thanks{E-mail: wangyanan@xao.ac.cn},
Y. J. Lei$^{4}$, G. Q. Ding$^{1}$, J. L. Qu$^{3}$, M. Y. Ge$^{3}$
\newauthor
C. M. Zhang$^{5}$, L. Chen$^{6}$, X. Ma$^{3}$ 
\\
$^{1}$Xinjiang Astronomical Observatory, Chinese Academy of Sciences,150, Science 1-Street, Urumqi, Xinjiang 830011, P.R. China\\
$^{2}$University of Chinese Academy of Sciences, 19A Yuquan road, Beijing 100049, P.R. China\\
$^{3}$Key Laboratory for Particle Astrophysics, Institute of High Energy Physics, Chinese Academy of Sciences, 19B Yuquan Road, Beijing 100049, P.R. China\\
$^{4}$Key Laboratory of Optical Astronomy, National Astronomical Observatories, Chinese
Academy of Sciences, Beijing 100012, P.R. China\\
$^{5}$National Astronomical Observatories, Chinese Academy of Sciences, Beijing 100012, P.R. China\\
$^{6}$Department of Astronomy, Beijing Normal University, Beijing 100875, P.R. China\\
}
\begin{document}

\date{Accepted ? December ?. Received ? December ?; in original form ? December ?}

\maketitle

\label{firstpage}

\begin{abstract}

  Using all the observations from {\it Rossi X-ray Timing Explorer} for the accreting neutron
star XTE~J1701-462, we carry out a systematic study on the cross-correlation function between
its soft and hard light curves.
During the entire outburst, XTE~J1701-462 evolves from super-Eddington luminosities
to quiescence, and shows both Z and atoll behaviours.
Following the previous work, we divide the outburst into five intervals,
one Cyg-like interval, three Sco-like intervals, and one atoll interval,
according to their different behaviors
in the corresponding color-color diagrams (CCDs).
With cross-correlation analyses, the anti-correlation, positive and ambiguous correlations are found in the different intervals in this source.
Both the anti-correlated soft and hard time lags are detected, where the hard lags
mean that the hard photons lag behind the soft ones and the soft lags are reverse.
In the Cyg-like interval,
the anti-correlations are presented in the hard vertex and upper
normal branch (NB), and the positive
correlations dominate the horizontal branch (HB) and lower NB. In the first two Sco-like intervals, the anti-correlations are firstly detected and most of them show up the HB and/or upper NB, and the positive correlations are mostly detected in the
lower NB and flaring branch (FB).
While in the following interval, i.e., the third Sco-like interval,
the anti-correlations occur on the upper FB, and the positive correlations mainly
distribute in the lower FB.
The different intervals are corresponding to the various luminosities,
therefore, the position of anti-correlations in the CCD might depend on the luminosity.
It is noted that,
in the Cyg-like interval, the positive correlations dominate
the HB, which is not consistent with that
of the Cyg-like Z sources GX~5-1 and Cyg~X-2 whose HBs host ambiguous correlations
and anti-correlations.
Hence, for comparing with GX~5-1,
we analyze the spectra of the HB and the hard vertex of the Cyg-like interval.
The fitting results show that, different from GX~5-1,
the ratio of the hard
emission to the soft emission basically keeps unvaried from the HB to the hard vertex,
which might result in the positive correlation.
Additionally, we compare the spectra of the third Sco-like interval with atoll
source 4U~1735-44, and find their spectral evolution along the tracks in the CCD is similar,
indicating that in
this interval, XTE~J1701-462 could approach an atoll source. The truncated disk model might be responsible for the detected time lags in XTE~J1701-462,
and the possible origins of the anti-correlated hard time lags and soft
time lags are also discussed.
\end{abstract}

\begin{keywords}
accretion, accretion disk--binaries: close--stars:
individual (XTE J1701-462)--X-rays: binaries
\end{keywords}
\section{Introduction}

Low-mass X-ray binaries (LMXBs) consist of a compact object that could be a black hole
candidate (BHC) or a neutron star (NS) and an evolving low mass star.
A compact object can accrete matter from the companion via its accretion disk.
Owing to their spectral and timing properties,
Hasinger \& van der Klis (1989) divided the NS LMXBs into
atoll and Z sources. In their
X-ray color-color diagrams (CCDs), Z sources with high luminosity
track out roughly the letter `` Z '' (i.e., Cyg-like Z sources)
or the Greek letter `` $\nu$ '' (i.e., Sco-like Z sources)
within the shorter timescales (hours to days), whereas atoll sources
with low luminosity show a C-shaped track within days to weeks or so.
In the CCDs of Z sources, from top to bottom, there are three main
parts which are the so-called horizontal branch (HB), normal branch (NB), and
flaring branch (FB) (Hasinger et al. 1990). The track of atoll sources in their
CCDs also shows spectral changes, from the
island to the banana state (van der Klis 2000; Altamirano et al. 2008).
While in this paper, corresponding to the work of Lin et al. (2009b),
atoll branches are also referred to as hard states (HSs) and soft states (SSs).

In briefly, the X-ray energy spectrum of LMXBs consists of two components,
which are thermal (soft) and non-thermal (hard) components. For NS LMXBs, it is
generally believed that the soft component originates from the accretion disk
or NS surface, while the hard component results from the Comptonization of soft photons
in the hot corona/Compton cloud. However, both the accretion disk structure and the location of the corona are not only complicated, but also unclear. The correlations and time lags between signals simultaneously detected in different energy bands, obtained with the cross-correlation function (CCF, Brinkman et al. 1974; Weisskopf et al. 1975), are helpful to explore the radiation mechanism and geometry of the accretion disk in different spectral states. The anti-correlation between the soft and hard light curves corresponds to negative cross-correlation coefficients (CCCs), and the positive CCCs are responsible for the positive correlation. The soft time lags mean that the higher energy photons lead lower energy ones, and vice versa, defined as the hard time lags. The detected time lags of X-rays in X-ray binaries span from
millisecond to kilosecond. The short scale ($\leq$ 1 s) time lags are usually attributed to the Comptonization of the soft photons from the accretion disk (Hasinger 1987; Nowak et al. 1999; B\"{o}ttcher \& Liang 1999). The long anti-correlated lags might be resulted from the viscous
timescale of the accretion disk, which is consistent with the truncated disk model.
Such anti-correlated time lags from dozens of seconds to thousands of seconds have been detected in the steep power law (SPL) state of some BH X-ray binaries (BHXBs) (e.g. GRS 1915+105, Choudhury et al. 2005; GX 339-4, Sriram et al. 2010), on the HB and upper NB of two Z sources (Cyg X-2, Lei et al. 2008; GX 5-1, Sriram et al. 2012), and on the upper banana (UB) state or the island state (IS) of two atoll sources (4U 1735-44, Lei et al. 2013; 4U 1608-52, Lei et al. 2014).

Based on the results of the cross-correlation analysis of different outbursts
in atoll source 4U 1608-52, Lei et al. (2014) suggested that the distribution of cross-correlation coefficient in the CCD might depend on the luminosity of the corresponding outburst.
However, similar studies for other sources have never been carried out.
Additionally, the evolution of cross-correlation has never been investigated in Sco-like
Z sources. XTE~J1701-462 could be a good candidate for such investigation, because it behaved as a  Cyg-like Z source at high luminosity, then with decreasing of luminosity, became a Sco-like Z source, and finally, switched into an atoll source at low luminosity (Homan et al. 2007a, c). Therefore, it is an ideal source for us to investigate the evolution of cross-correlation between Z sources and atoll sources and between Cyg-like Z sources and Sco-like Z sources as well.

XTE~J1701-462 was found on January 18 of 2006 by  {\it Rossi X-ray Timing Explorer} ({\it RXTE}) (Remillard et al. 2006). Afterwards, Homan et al. (2006) found that it showed behaviors of Z sources. Generally,  it is suggested that the mass accretion rate ($\dot{M}$) monotonically increases from the HB, via the NB, to the FB (Priedhorsky et al. 1986; Hasinger et al. 1990). However, based on their spectral analyses, Lin et al. (2009b) proposed that the motion along the Z branches in this source might be driven by three different physical mechanisms with roughly constant $\dot{M}$. Homan et al. (2007c) and Sanna et al. (2010) analyzed its Low frequency quasi-periodic oscillations (LFQPOs) and kilo-hertz QPOs (kHz QPOs), respectively. The type I X-ray bursts were observed (Lin et al. 2009a). Fridriksson et al. (2010) suggested that the NS crust of XTE~J1701-462 was rapidly cooling during its quiescent phase. Ding et al. (2011) proposed that it hosts a radiation-pressure-dominated accretion disk and its NS surface magnetic field strength is $\sim$(1--3)$\times10^9$~G. Li et al. (2013) analyzed its short scale time lags with the Fast Fourier Transform (FFT) algorithm and detected hard time lags and soft time lags. The radio observations for this peculiar source were also made with the Australia Telescope Compact Array (ATCA) on 22, January of 2006 (Fender et al. 2006).

In this work, using all the data on board {\it RXTE} during the 2006-2007 outburst of XTE~J1701-462, we systematically analyze the cross-correlation and long scale time lags of this source. Additionally, we compare its spectra with those of other NS LMXBs. We describe the observations and data reduction in Section 2, report the results in Section 3, present the summary and discussion in Section 4, and give our conclusions in Section 5.

\section{OBSERVATIONS AND DATA REDUCTION}

We analyze all the available data of XTE~J1701-462 on board {\it RXTE}, which consists of three instruments, the Proportional Counter Array (PCA; Jahoda et al. 2006), the High Energy X-ray
Timing Experiment (HEXTE; Gruber et al. 1996), and an All-Sky Monitor
(ASM; Levine et al. 1996). There are 5 non-imaging, collimated Xenon/methane
multi-anode proportional counter units (PCUs) of PCA, of which the best-calibrated and
the longest observational duration unit, PCU 2, has been selected. Following Lei et al. (2008), we use the standard2 mode data from the observations with exposure time lasting longer than 2000s to generate light curves. Using {\it PCABACKEST} V3.8, we produce PCA background files to extract the background light curves. When the source intensity is lower than 40 counts s$^{-1}$ PCU$^{-1}$, the faint background model is used, otherwise the bright background model is applied. Thus, we obtain the background-subtracted light curves in soft energy band (2-3 keV) and hard energy band (12-30 keV). We filter out those data with the following criteria: time since the peak of the last South Atlantic Anomaly
passage longer than 30 min, Earth elevation angle greater than 10\degr, the spacecraft pointing
offset less than 0.02\degr, and electron contamination less than 0.1. In addition, the 150 s data before a breakdown event and 600 s data after it, together with the data during the breakdown event, are excluded.

Based on the evolution properties of XTE~J1701-462, Lin et al. (2009b) divided the whole
outburst into five intervals: interval \textrm{I} belonging to a
Cyg-like Z source; intervals \textrm{II-III} being Sco-like Z sources; interval \textrm{IV} being regarded as the Z source FB stage, but, during which the source resembles some bright atoll sources somehow; interval \textrm{V}, a characteristic atoll source. Figure 1 shows the PCA light curve of the observations with segments lasting longer than 2000 s.

\begin{figure}
\centering
\includegraphics[width=7cm]{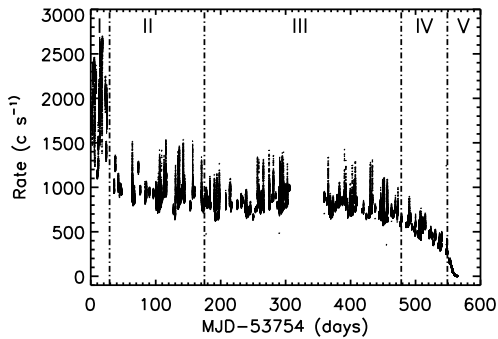}
\caption{The {\it RXTE} PCA light curve with segments longer than 2000 s
during the 2006-2007 outburst. According to Lin et al. (2009b), the outburst is divided
into five intervals, signed by I-V.} \label{fig1}
\end{figure}

We use the {\it XRONOS} tool ``{\it crosscor}'' to calculate the cross-correlation between the soft (2-3 keV) and hard (12-30 keV) light curves. Liking the assorting of the cross-correlation of Z source Cyg X-2 (Lei et al. 2008), the results of CCF are classified into positive, ambiguous, and anti-correlations. In order to obtain CCCs and time lags, following Sriram et al. (2007), we fit the peak of CCF using a Gaussian function, with a 90\% confidence level. If an observation includes several segments, each of which lasts longer than 2000 s and shows obvious correlation, following Sriram et al. (2012), we calculate the cross-correlation of each segment. In order to obtain the statistical distribution of CCCs, we define a positive correlation as the CCC to be higher than 0.1 and an anti-correlation as the CCC to be lower than -0.1. The remains are ambiguous correlations. The background-subtracted soft and hard light curves
of four representative observations with anti-correlation, and their CCFs
are shown in the left and middle columns of Figure 2, respectively. In order to study spectral change, we produce the background-subtracted spectra of the hard region (or soft region) where the count rate ratio of 12-30 keV to 2-3 keV is larger than 1.1 times (or less than 0.9 times) the average count rate ratio of the two energy bands. Then, we calculate the spectral ratios of the hard region spectra to the soft region spectra. The hard region spectra, soft region spectra, and spectral ratios of the four observations are demonstrated in the right column of Figure 2.
As shown by the low panels of the four panels of the right column of Figure 2, with increasing energy, the spectral ratio increases and the ratio line goes across the solid transverse line with unit spectral ratio. There exists a spectral pivoting energy, i.e the energy of the unit spectral ratio. In the energy bands higher than the spectral pivoting energy, the source intensity is relatively larger in the hard region than in the soft region, while in the energy bands lower than the spectral pivoting energy, the source intensity is relatively larger in the soft region than in the hard region (Shirey et al. 1999). The spectral pivoting confirms the anti-correlation. We fit the ratio line around the spectral pivoting energy with a straight line function and get the spectral pivoting energy and its errors, which are deduced from the error transfer function.
The positive and ambiguous correlations are shown in Figures 3 and 4, respectively.

\begin{figure*}
\centering
\includegraphics[width=6cm]{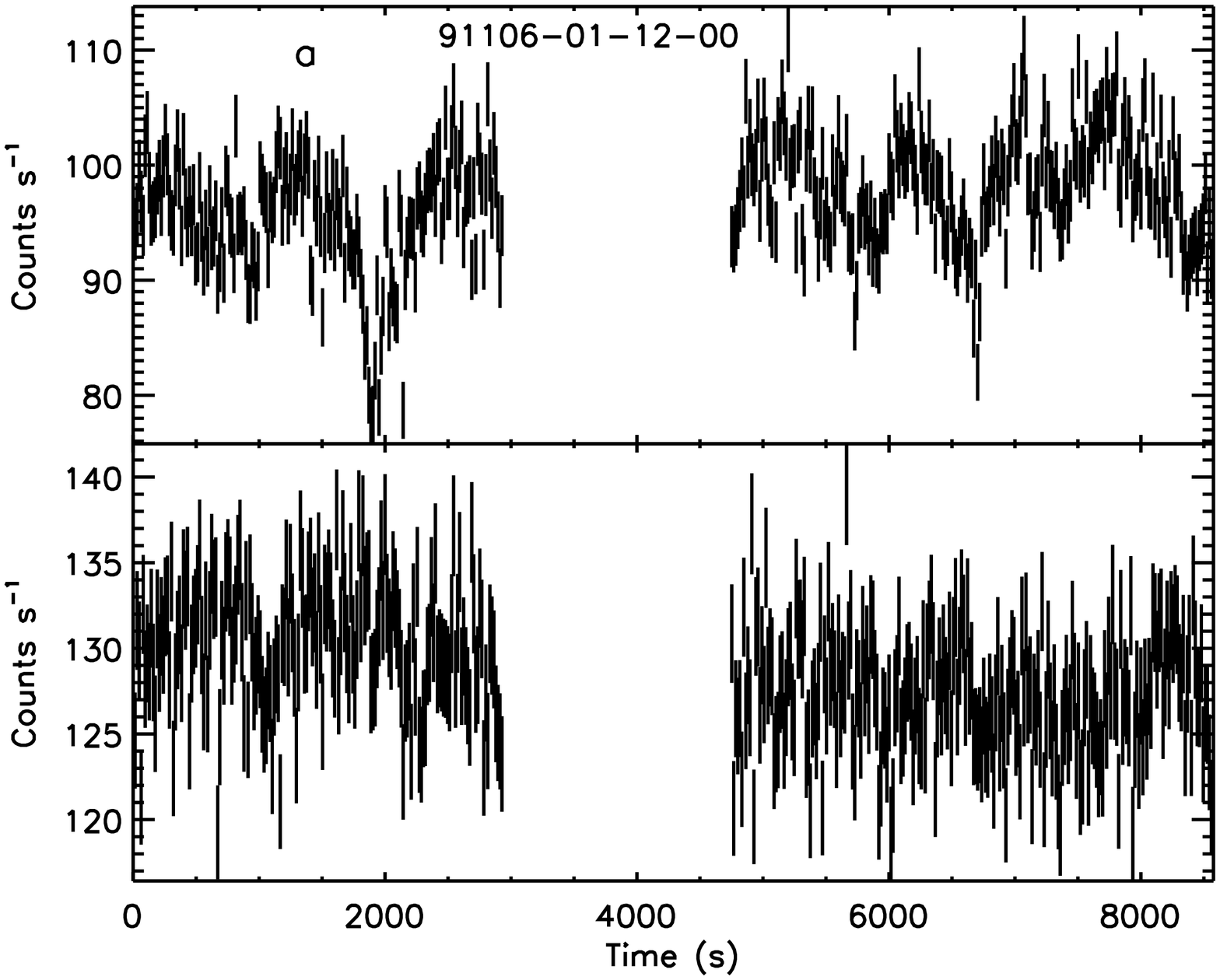}
\hspace{-2.2em}
\includegraphics[width=6cm]{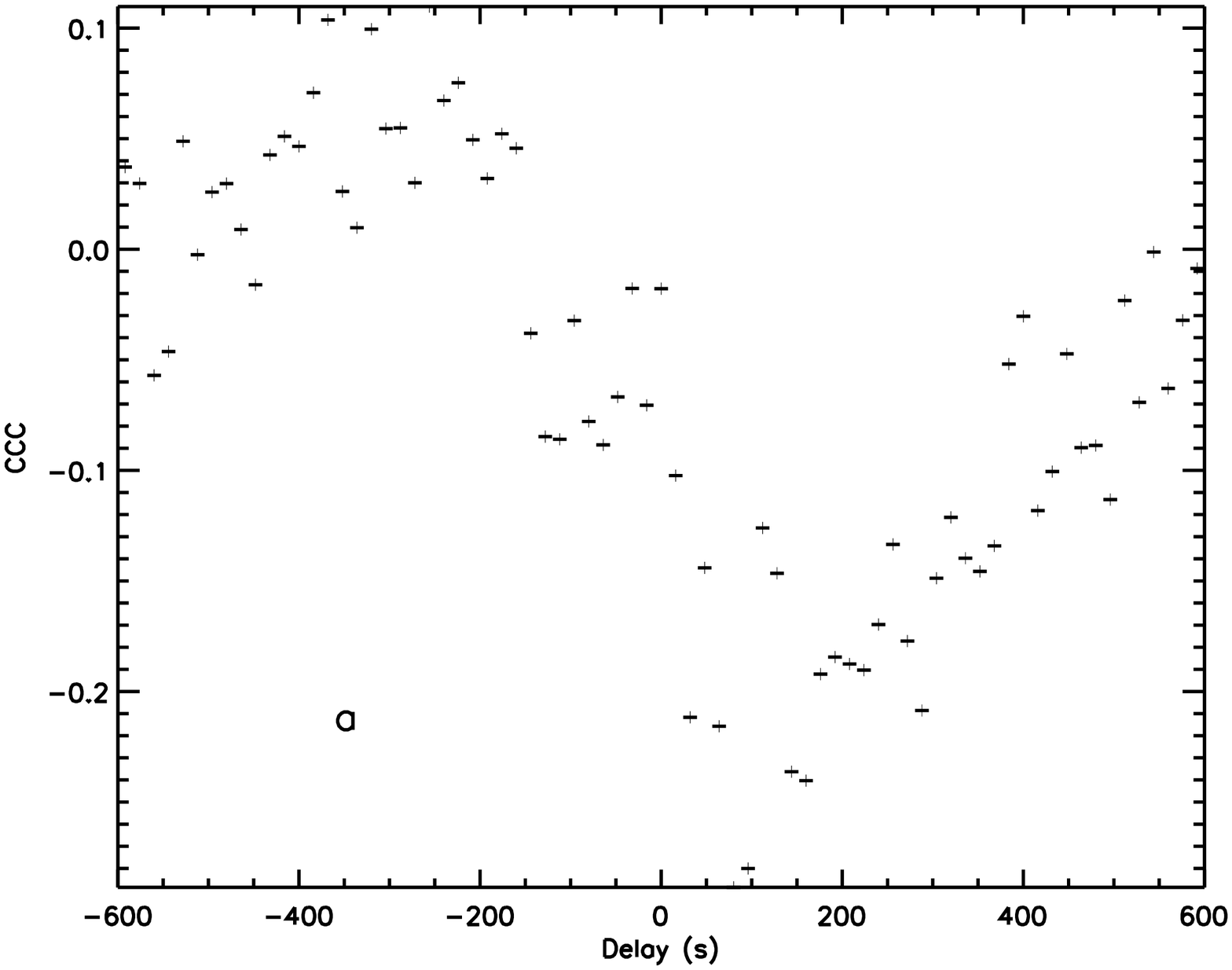}
\hspace{-2.2em}
\includegraphics[width=6cm]{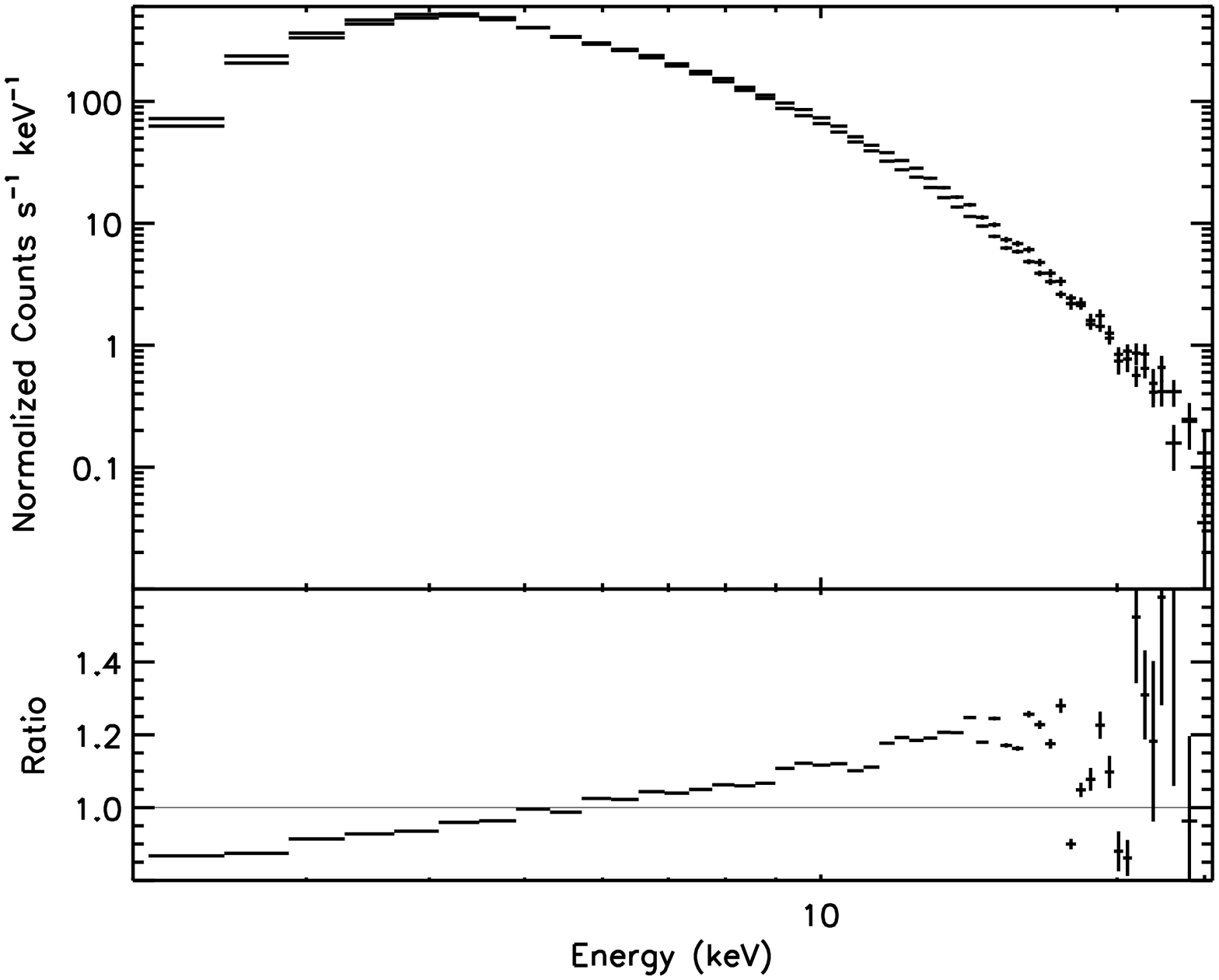}
\hspace{-2.2em}
\includegraphics[width=6cm]{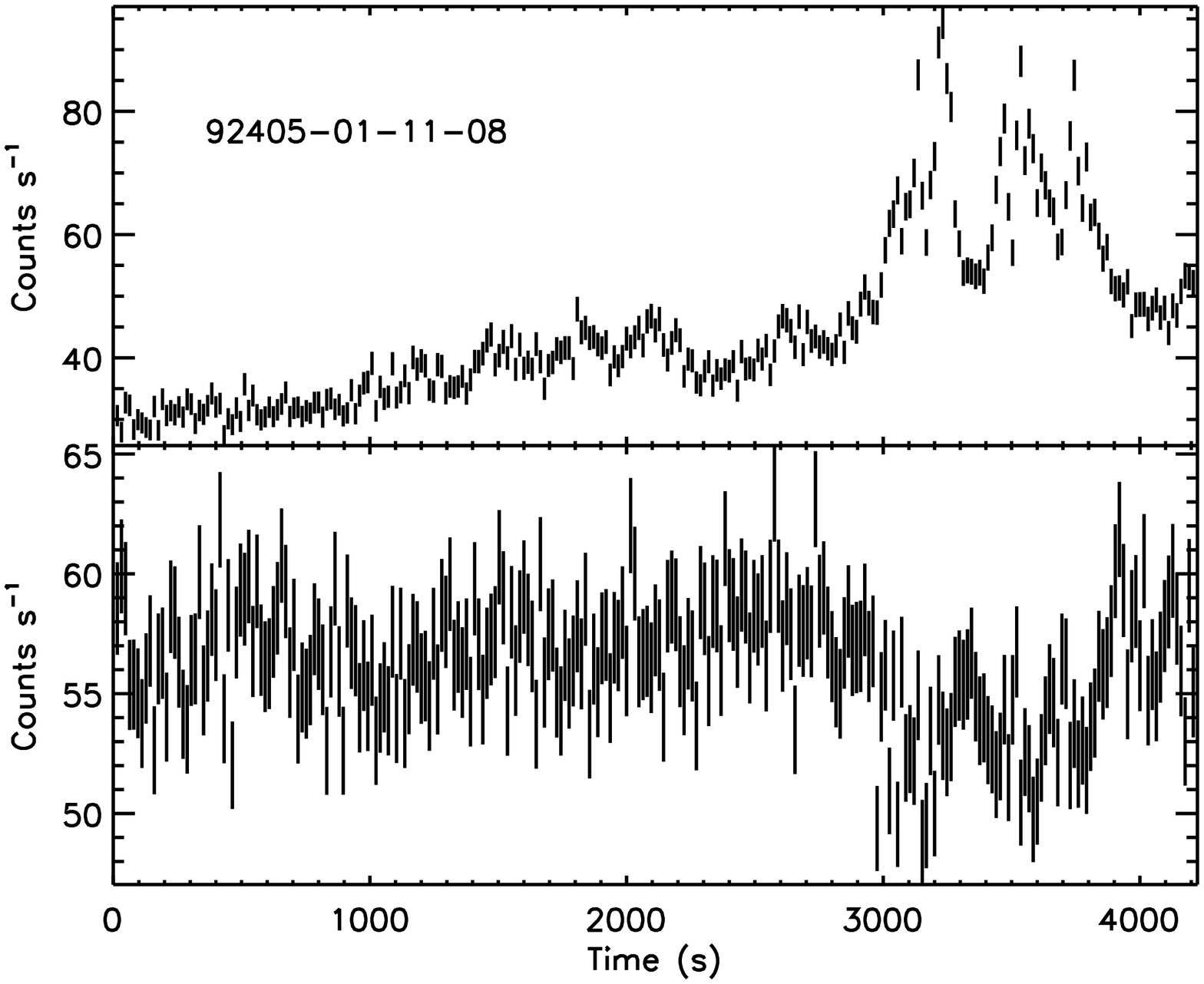}
\hspace{-2.2em}
\includegraphics[width=6cm]{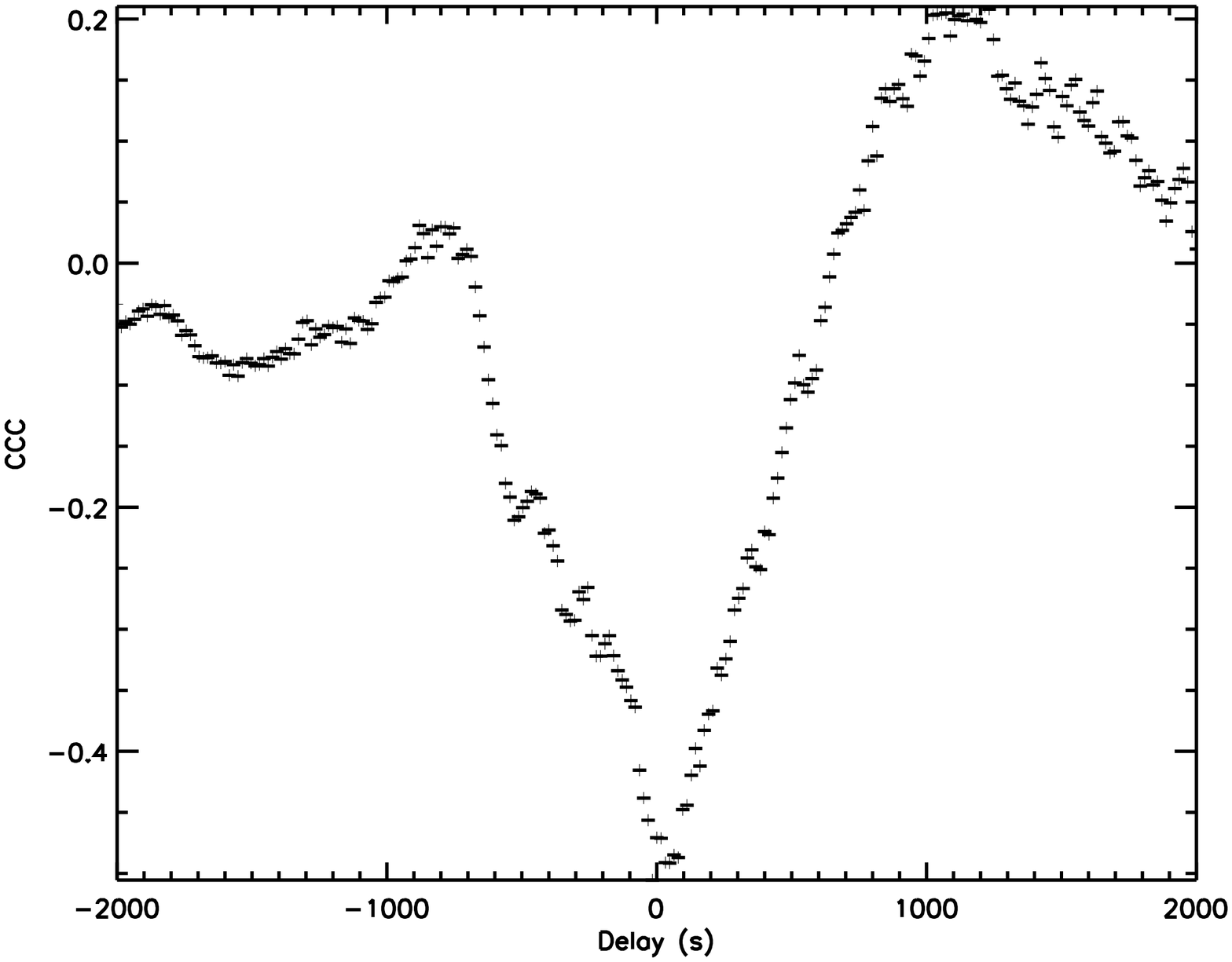}
\hspace{-2.2em}
\includegraphics[width=6cm]{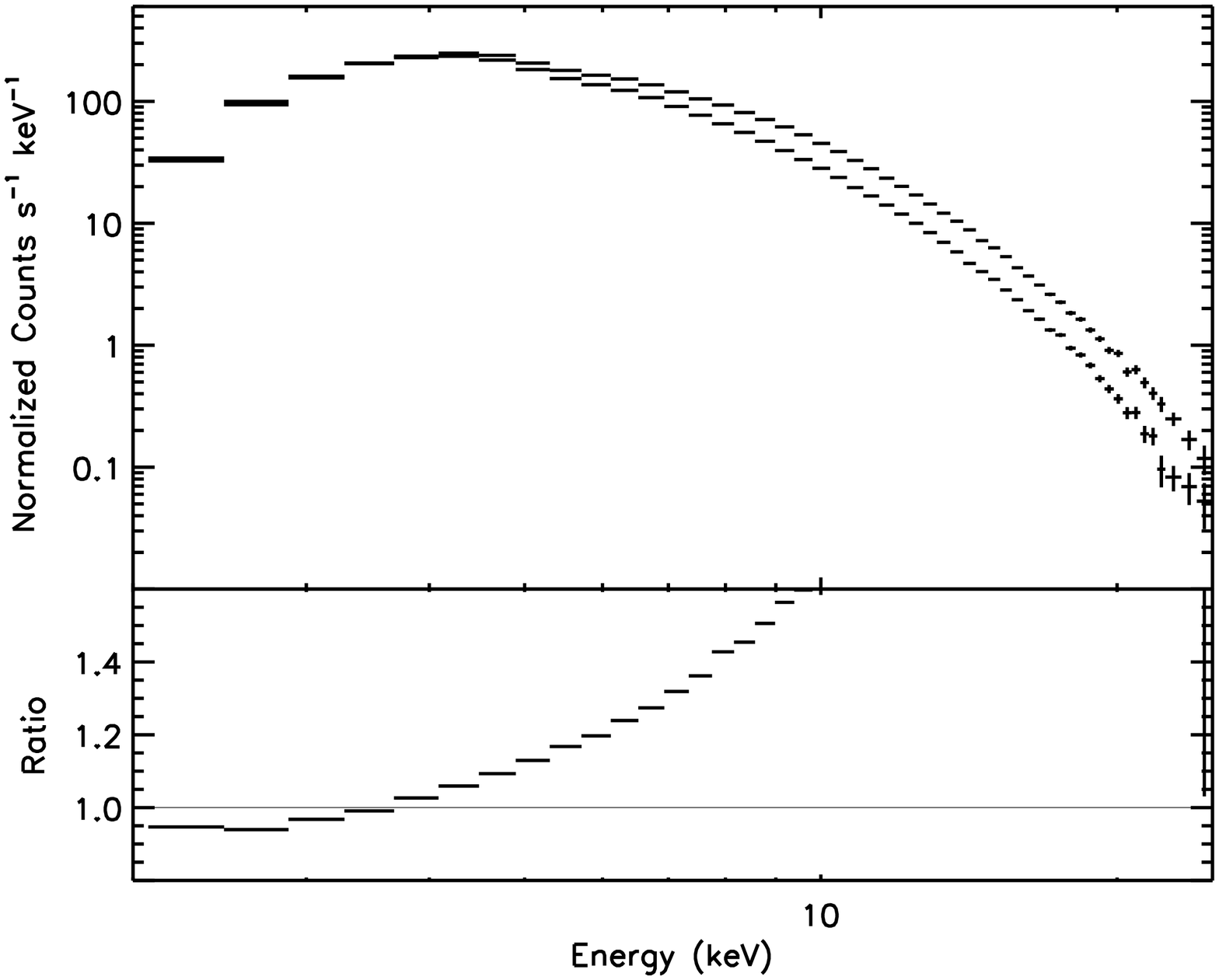}
\hspace{-2.2em}
\includegraphics[width=6cm]{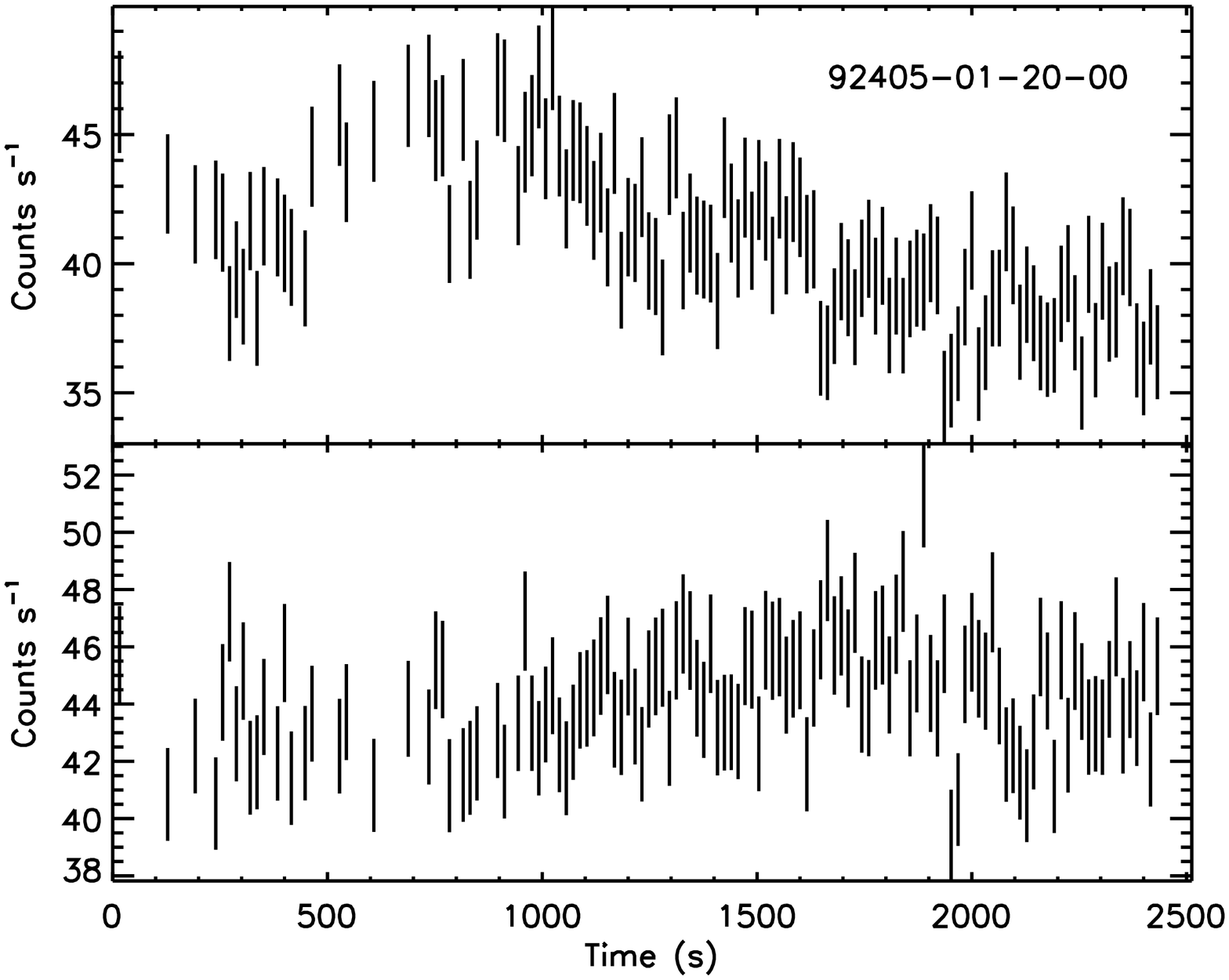}
\hspace{-2.2em}
\includegraphics[width=6cm]{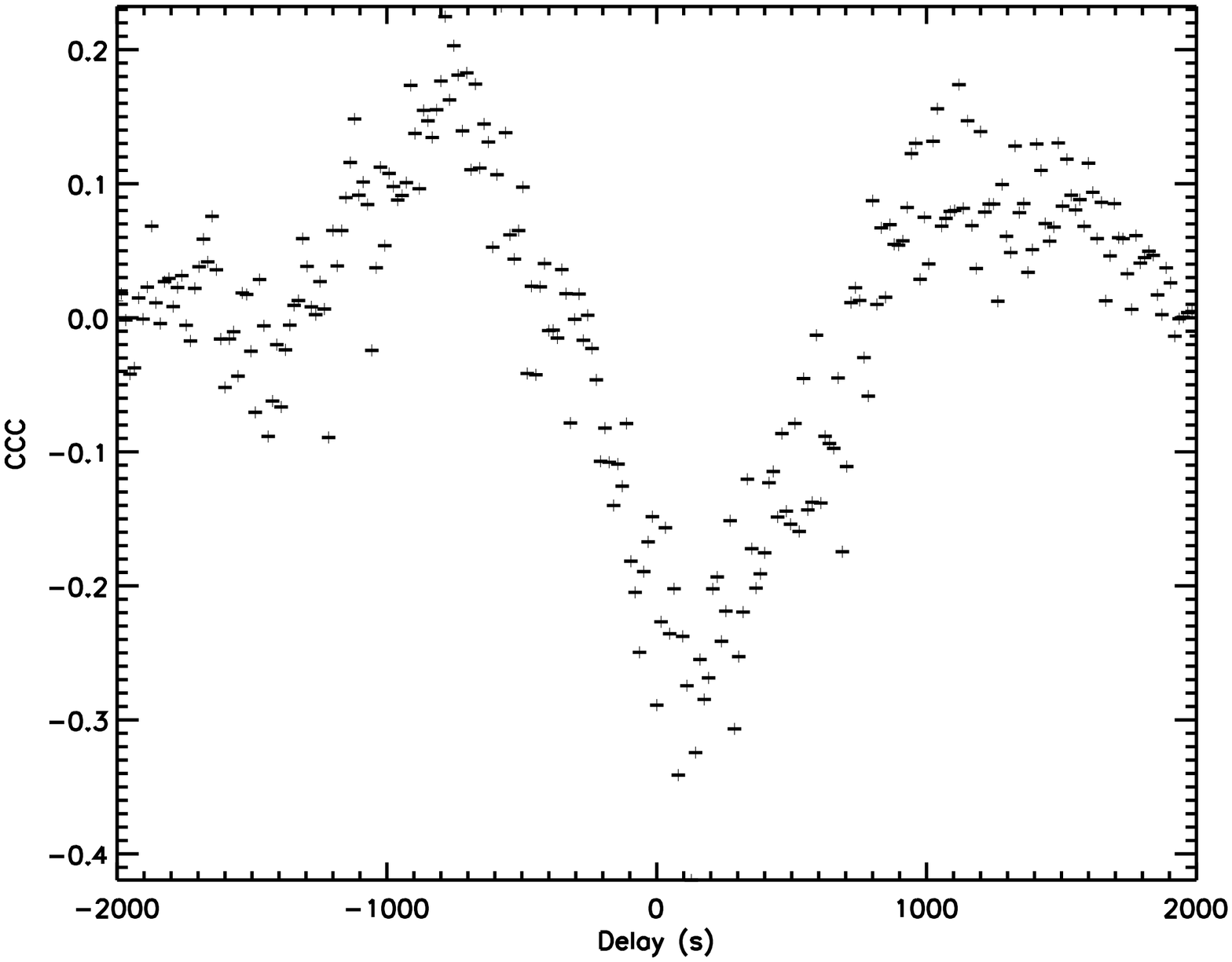}
\hspace{-2.2em}
\includegraphics[width=6cm]{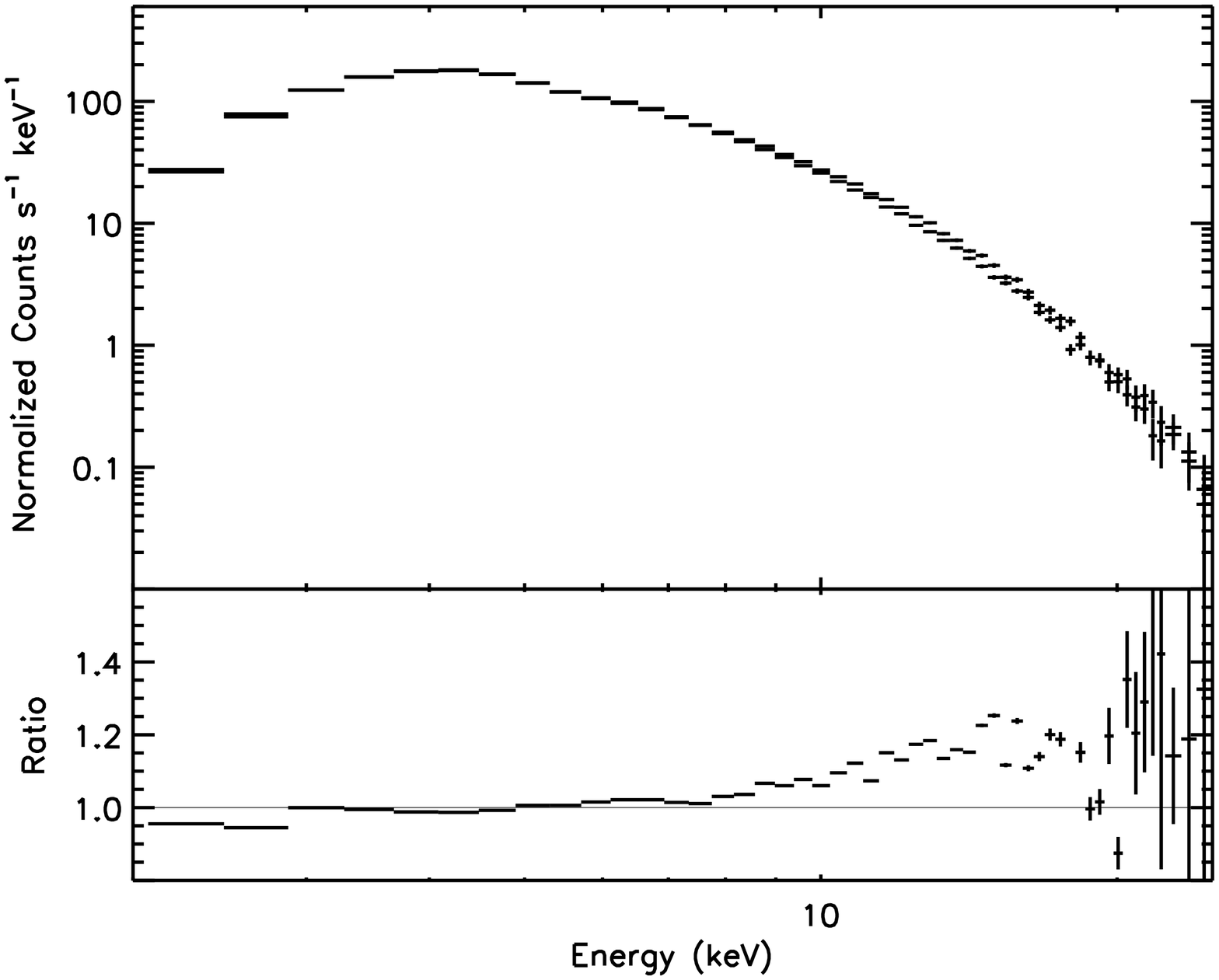}
\hspace{-2.2em}
\includegraphics[width=6cm]{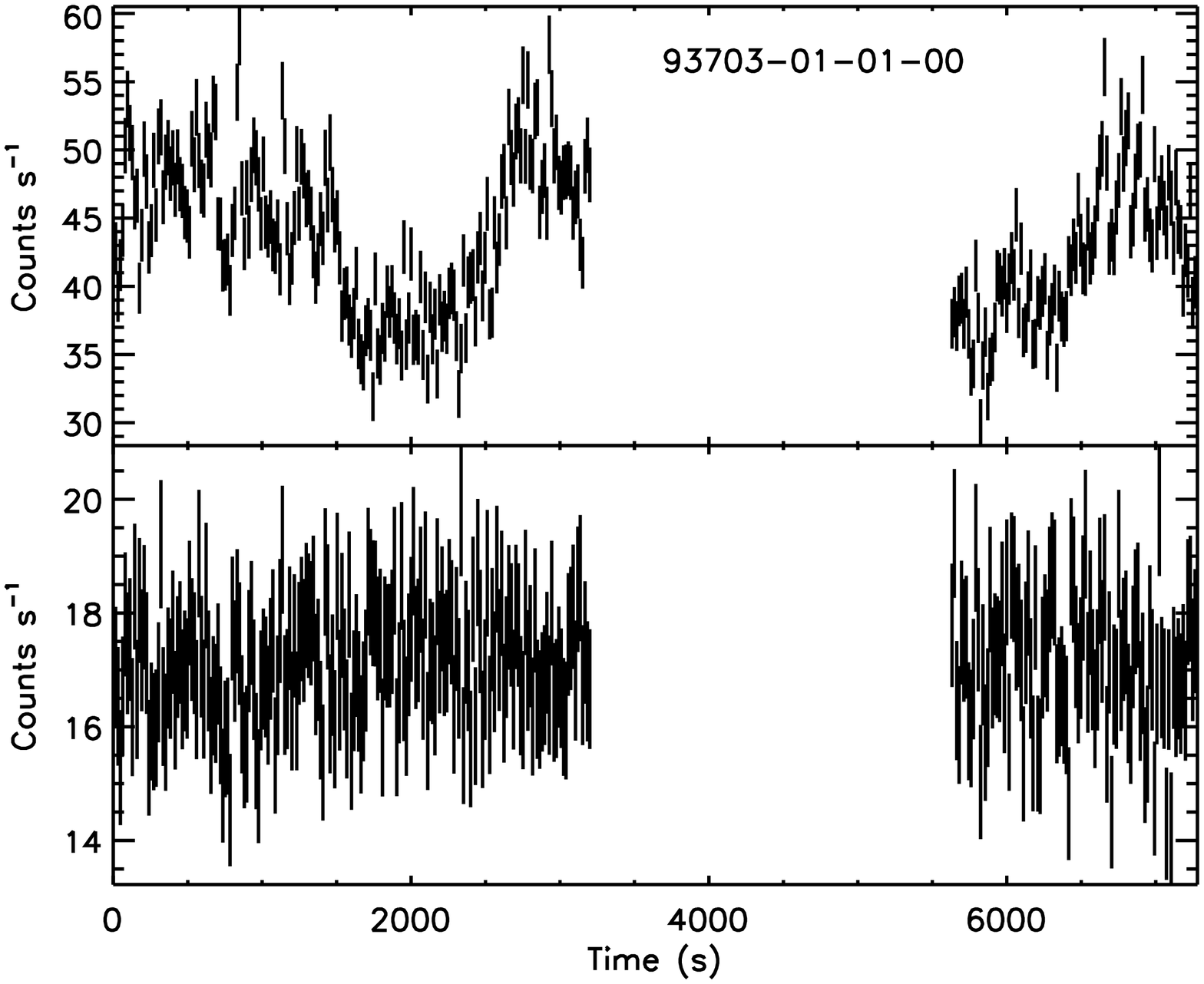}
\hspace{-2.2em}
\includegraphics[width=6cm]{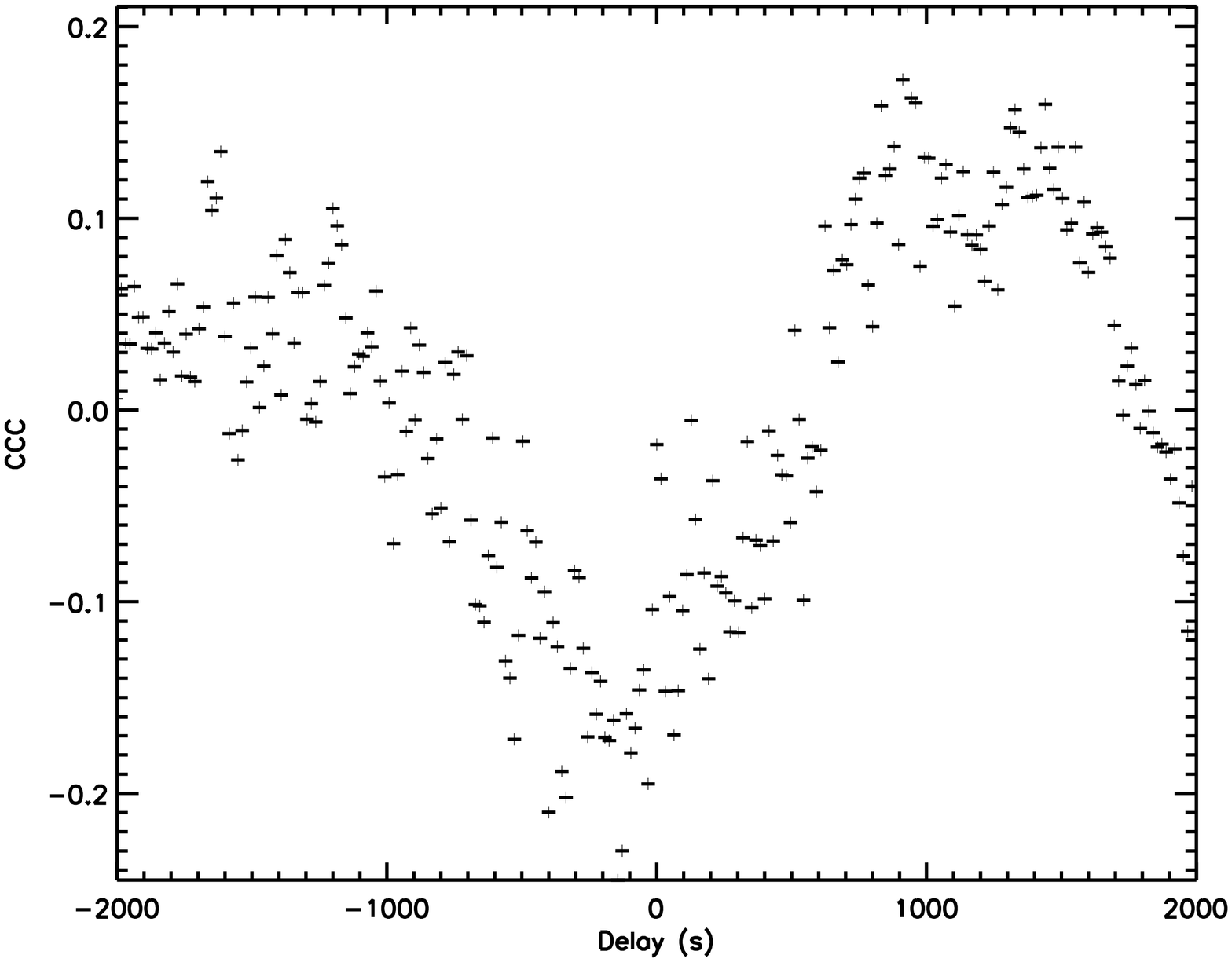}
\hspace{-2.2em}
\includegraphics[width=6cm]{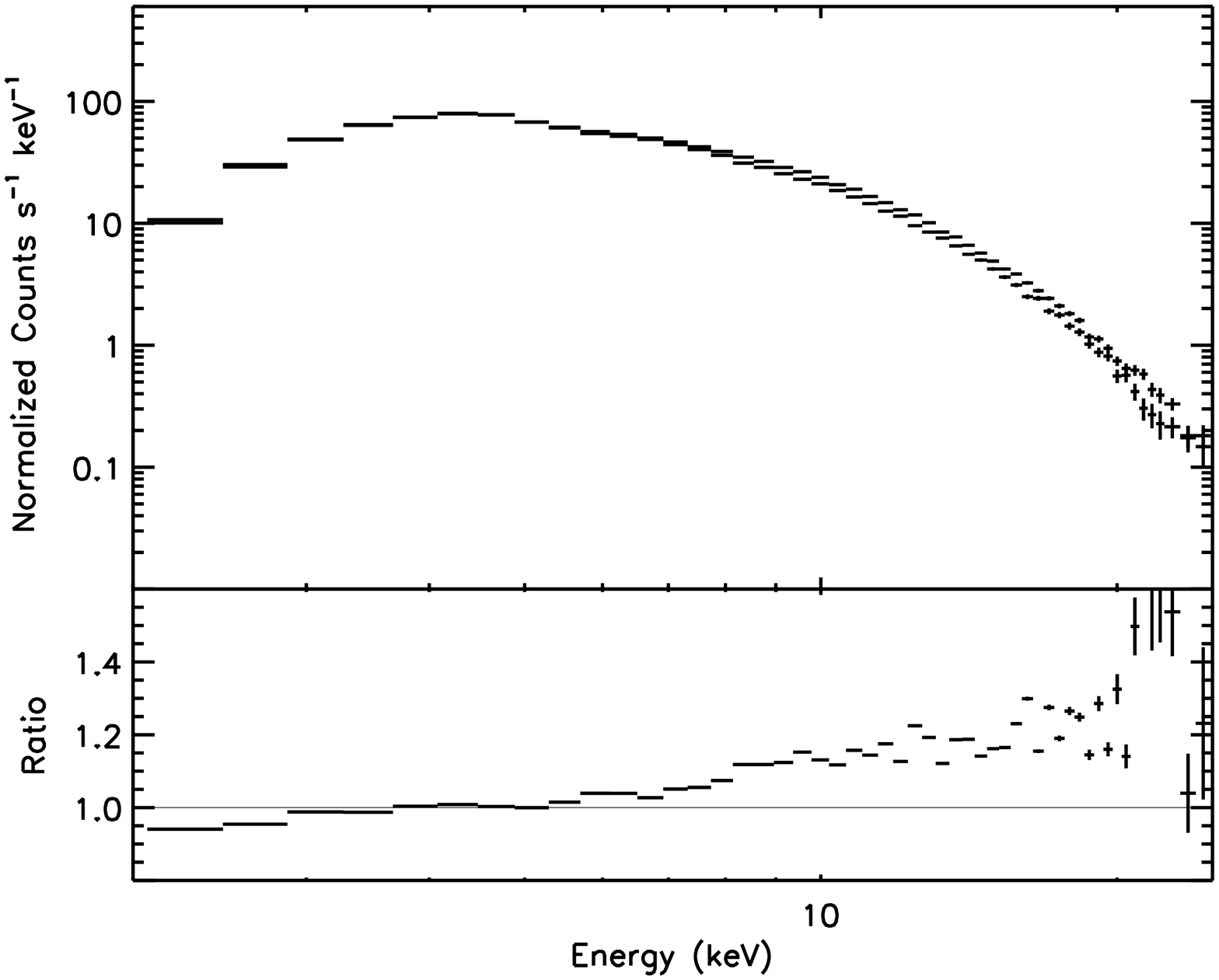}
\caption{Left column: the hard light curves (upper panels, 12-30 keV) and soft light curves (low panels, 2-3 keV) of four representative observations with anti-correlations; middle column: the cross-correlation functions of the four observations; right column: the hard region spectra and soft region spectra and their ratios of the four observations. In the upper panel of ObsID 91106-01-12-00 in left column, the anti-correlation is detected from the segment labeled by the letter `a', which is longer than 2000 s.}
\end{figure*}

\begin{figure*}
\centering
\includegraphics[width=6cm]{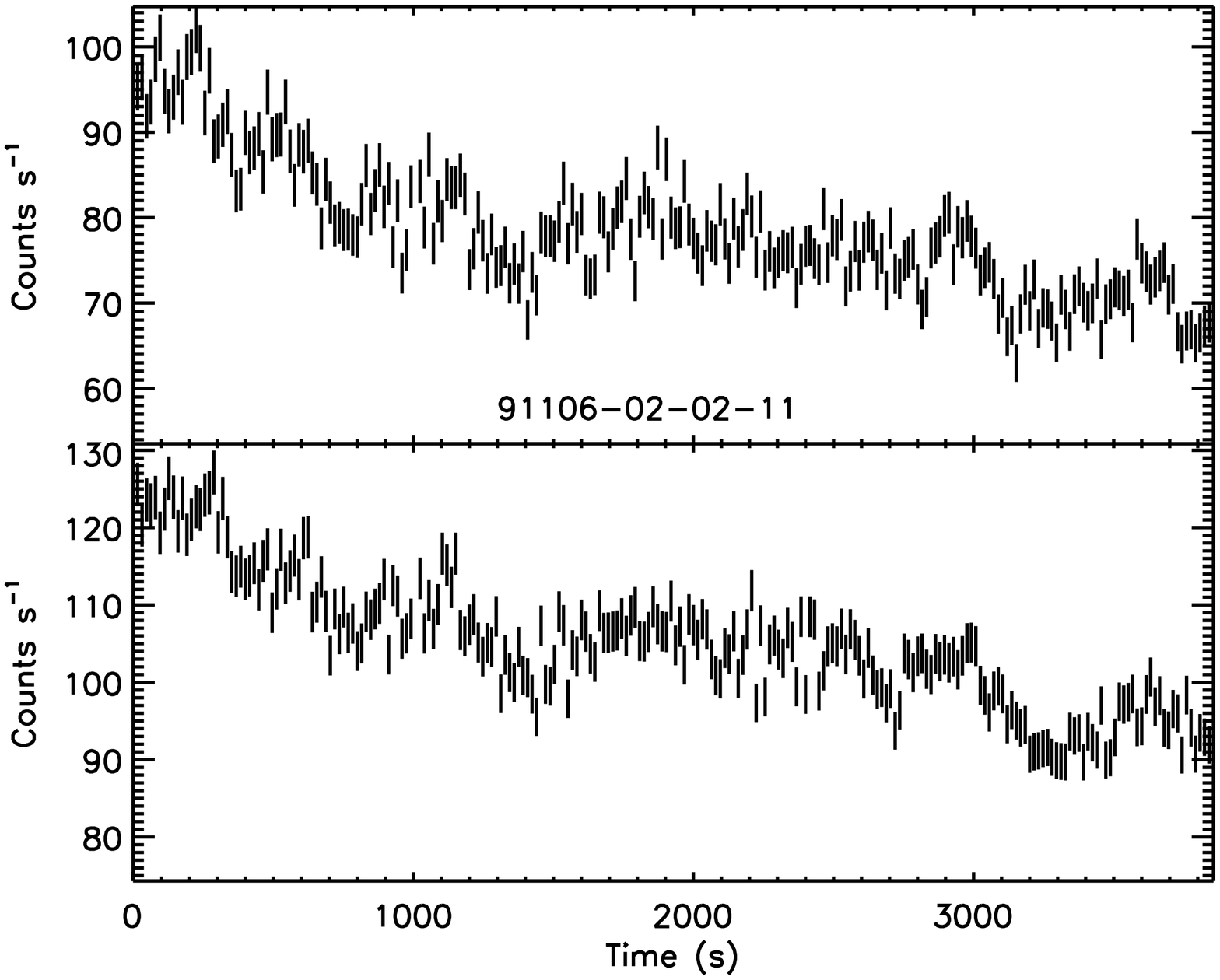}
\hspace{-2.2em}
\includegraphics[width=6cm]{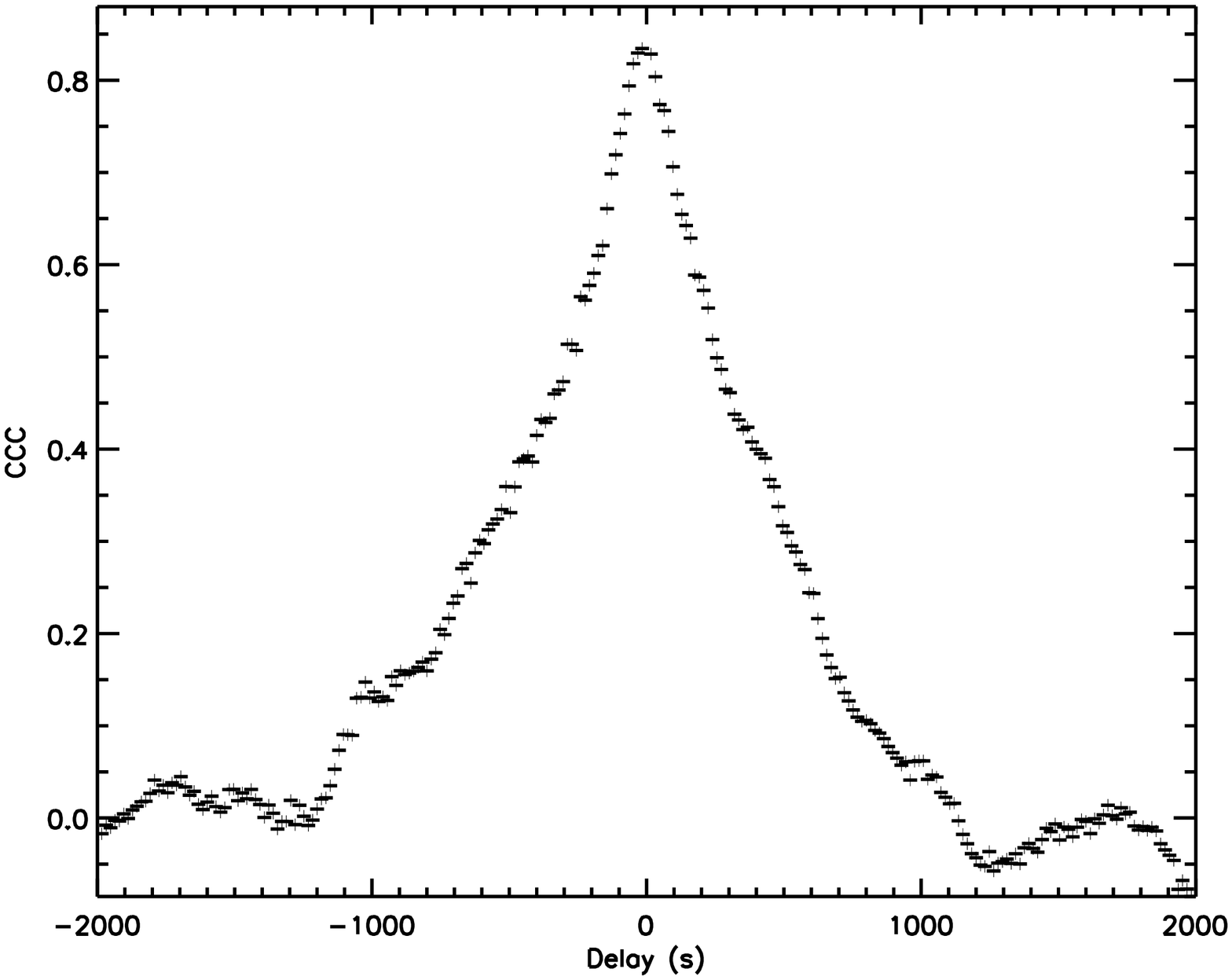}
\hspace{-2.2em}
\includegraphics[width=6cm]{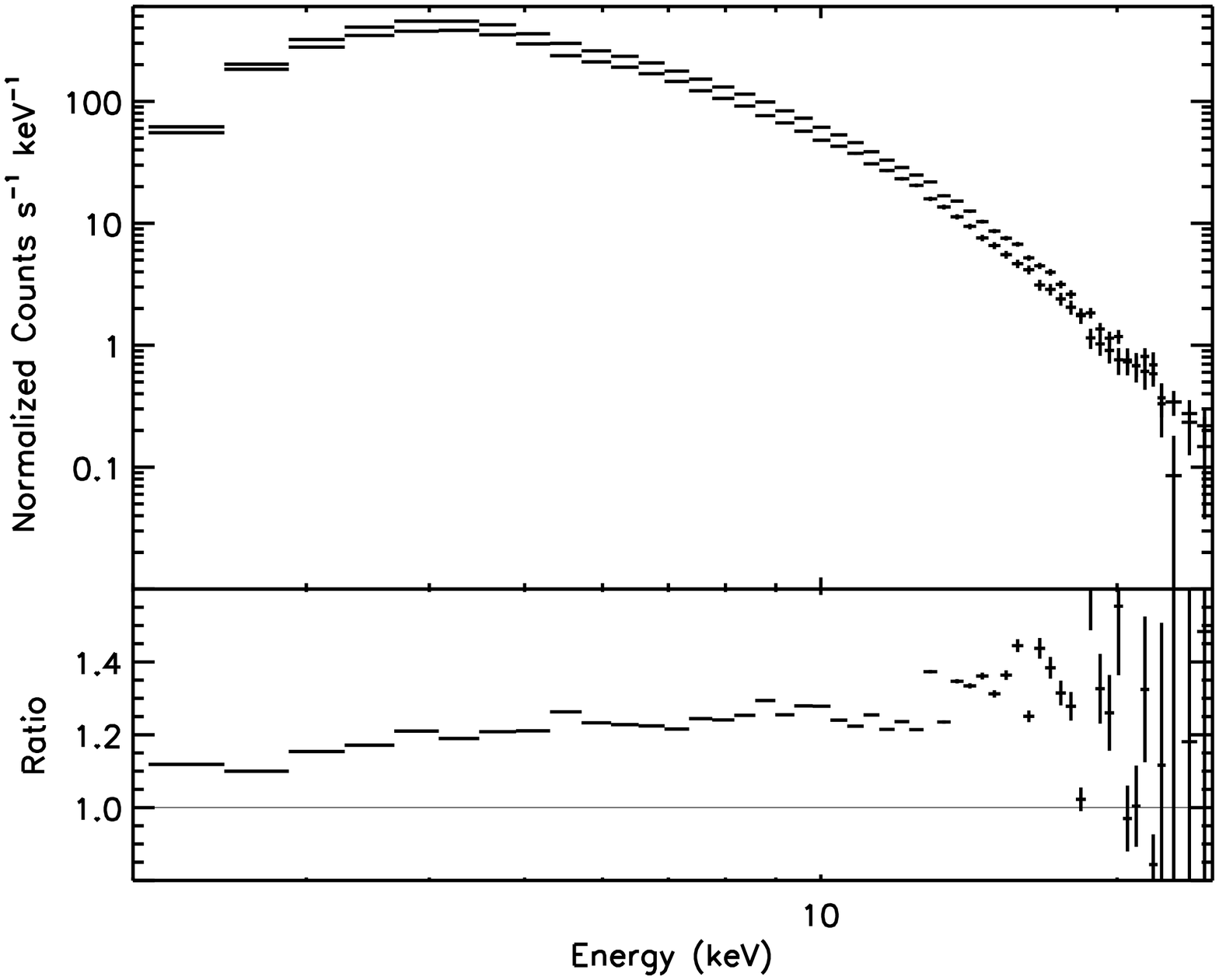}
\caption{The upper panel and low panel on the left respectively show the hard light curve (12-30 keV) and soft light curve (2-3 keV) of a representative observation with positive correlation; the middle panel: the cross-correlation function of the observation; the hard region spectrum and soft region spectrum of the observation and their ratios are separately demonstrated by the upper panel and low panel on the right.}
\end{figure*}

\begin{figure*}
\centering
\includegraphics[width=6cm]{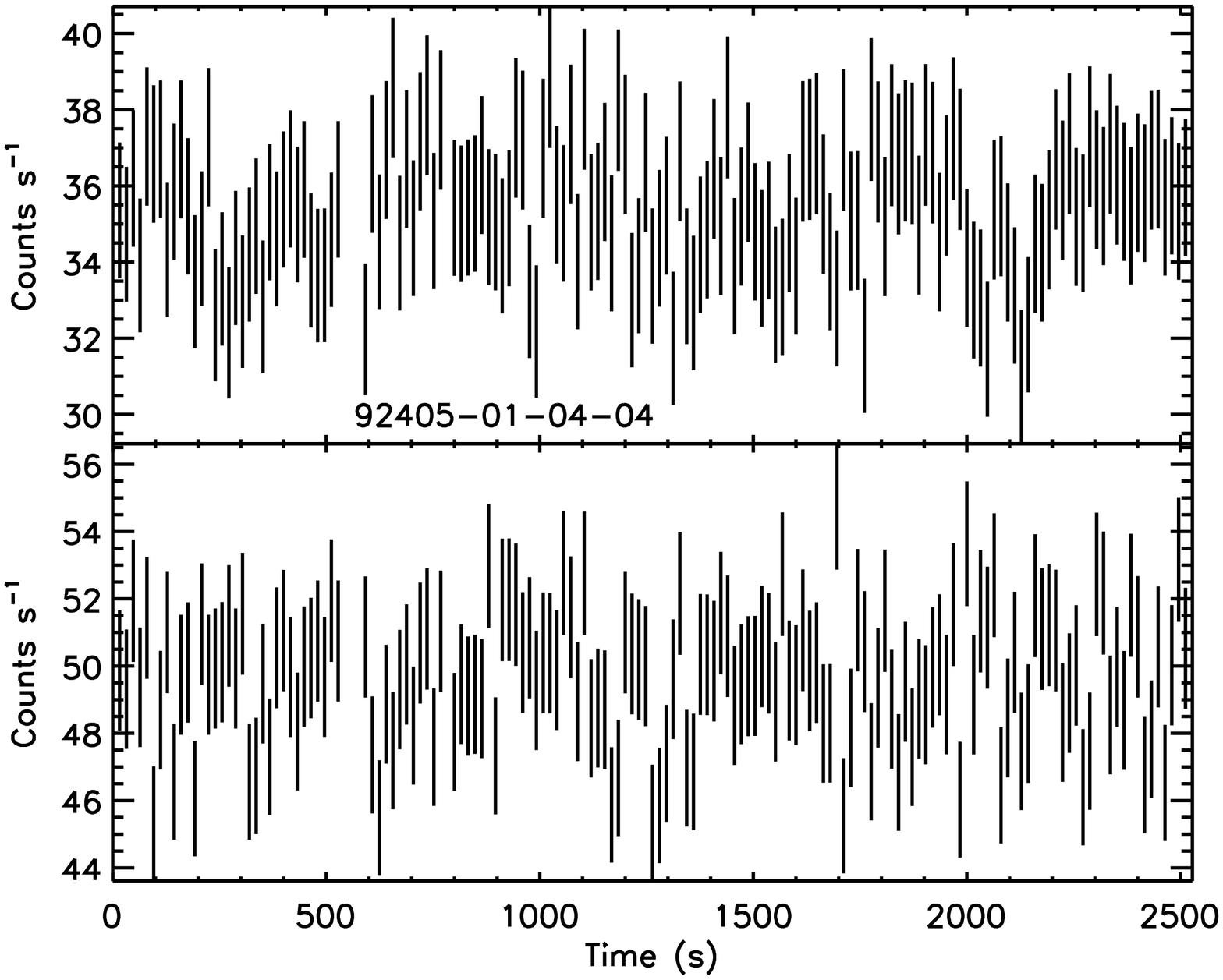}
\hspace{-2.2em}
\includegraphics[width=6cm]{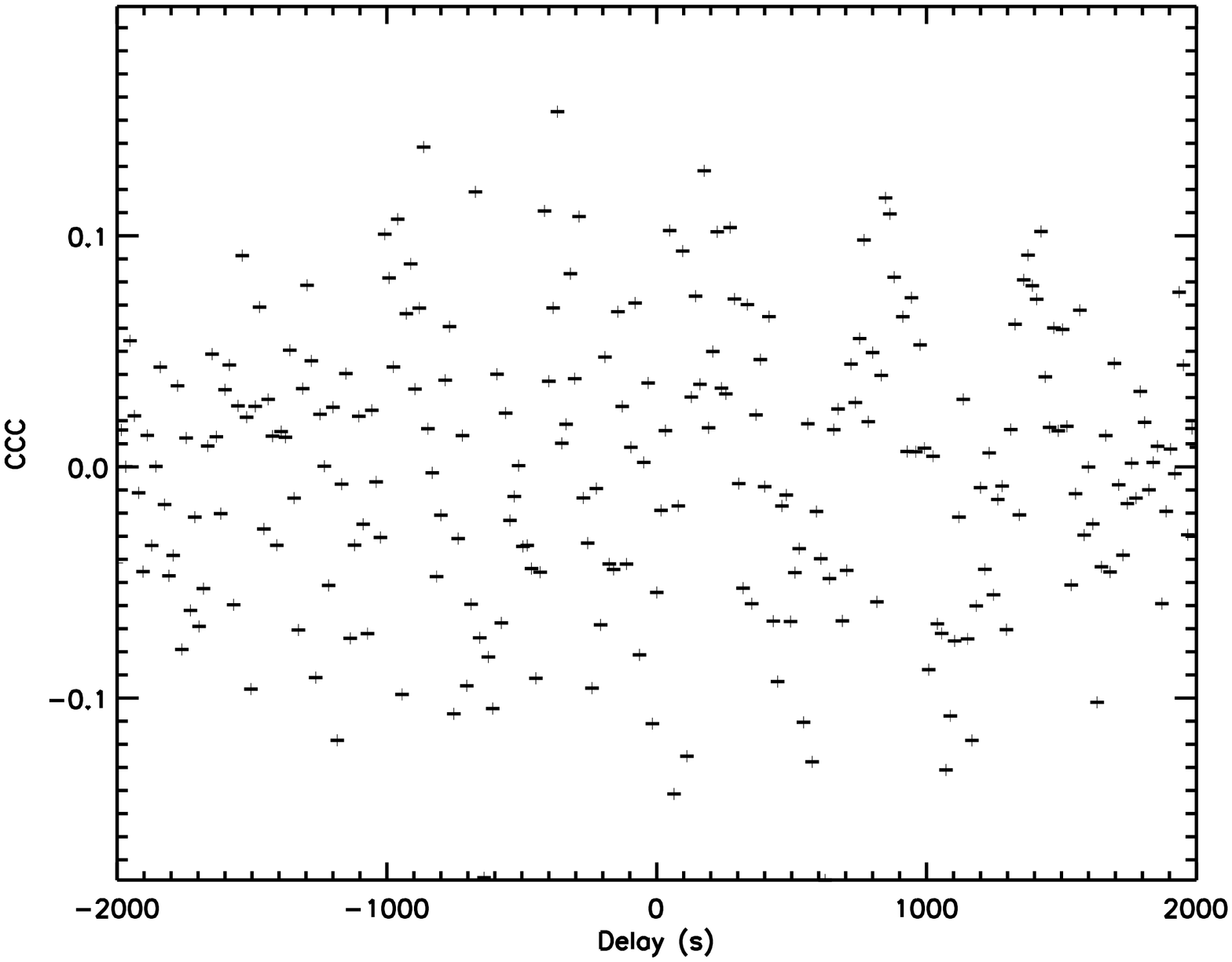}
\hspace{-2.2em}
\includegraphics[width=6cm]{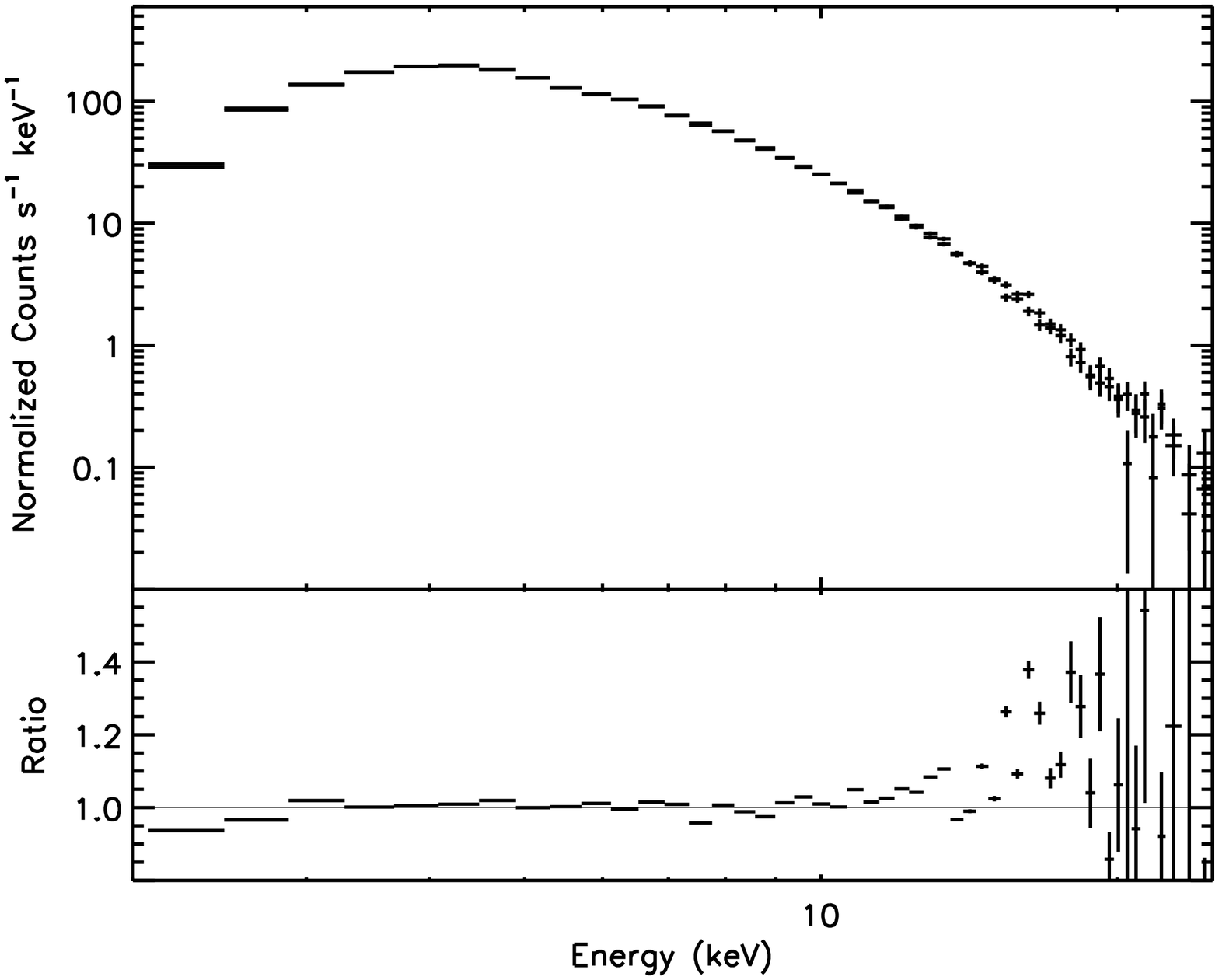}
\caption{The same as in Figure 3, but the representative observation with ambiguous correlation.}
\end{figure*}

Figure 5 shows the evolution of the CCDs of XTE~J1701-462, where
soft and hard colors are defined as
4.5-7.4 keV/2.9-4.1 keV and 10.2-18.1 keV/7.8-9.8 keV count rate ratios,
respectively (Li et al. 2013). In order to minimize
the statistical uncertainties, the CCDs are produced from 1600 s bin (except
interval V, each point representing one PCA observation). The branches of each CCD are marked.
The PCA spectra are fit with the spectral models using {\it XSPEC} V12.7.1.
Due to the calibration uncertainties of low-energy channels,
a 0.6\% systematic error is added to the spectra.
The {\it RXTE} spectra and simultaneous {\it Swift} spectra are jointly fit and the interstellar absorption hydrogen column density {\it N$_{H}$}
is fixed at $2.0 \times 10^{22}$ cm$^{-2}$ (Lin et al. 2009b).

\begin{figure}
%\begin{center}
\vspace{-1.5em}
\includegraphics[width=7cm]{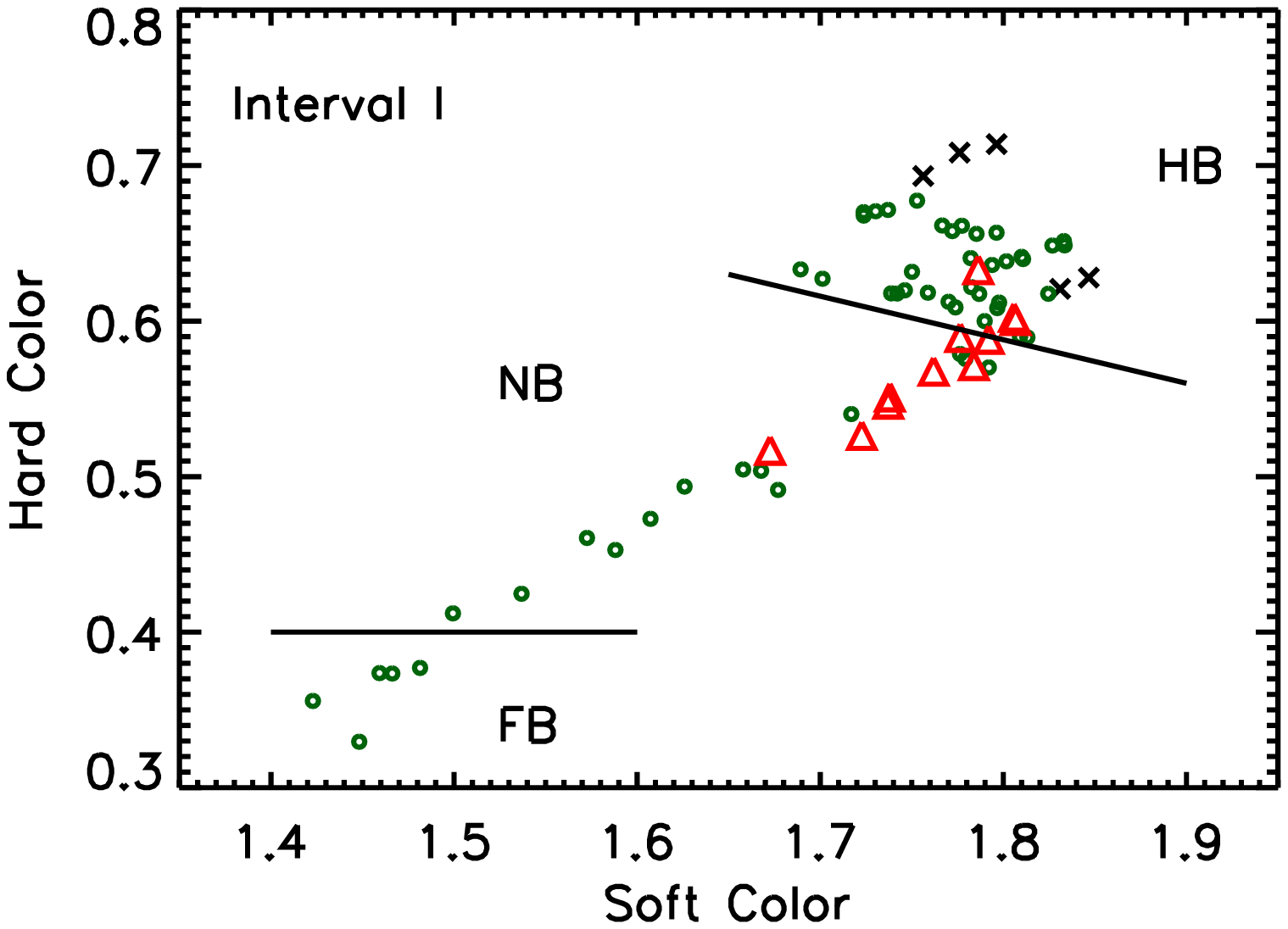}
\vspace{-2.0em}
\includegraphics[width=7cm]{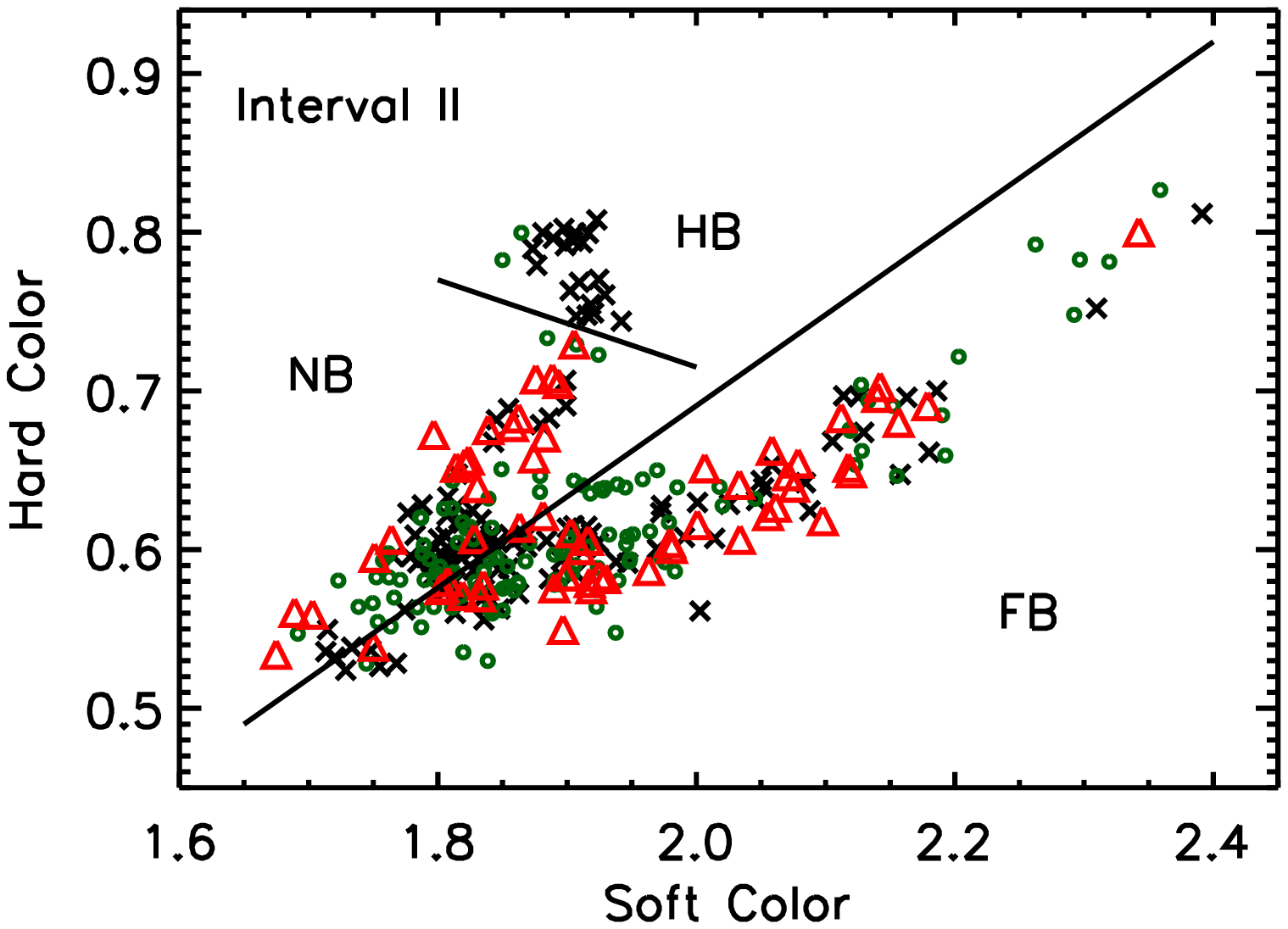}
\vspace{-2.0em}
\includegraphics[width=7cm]{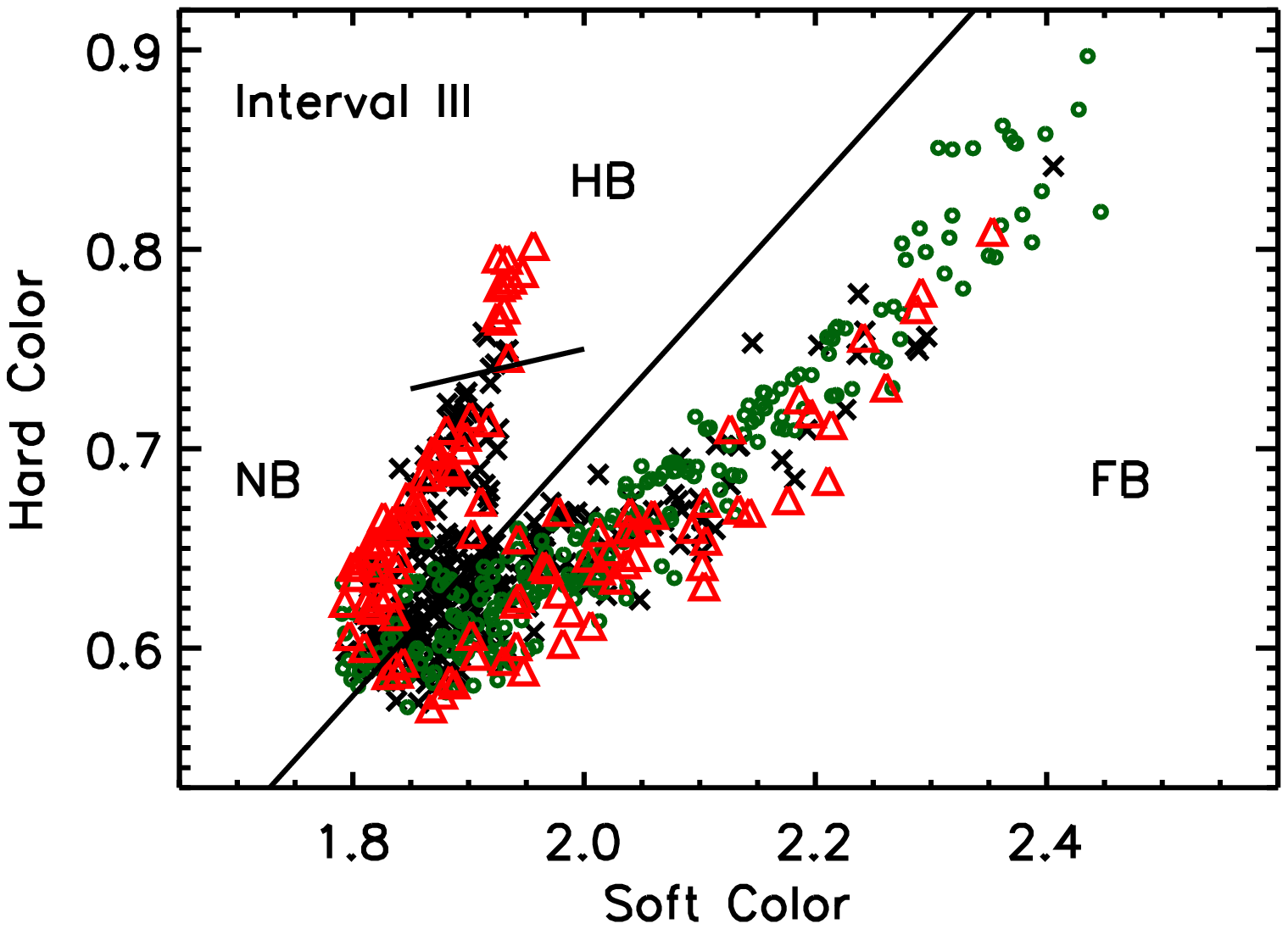}
\vspace{-2.0em}
\includegraphics[width=7cm]{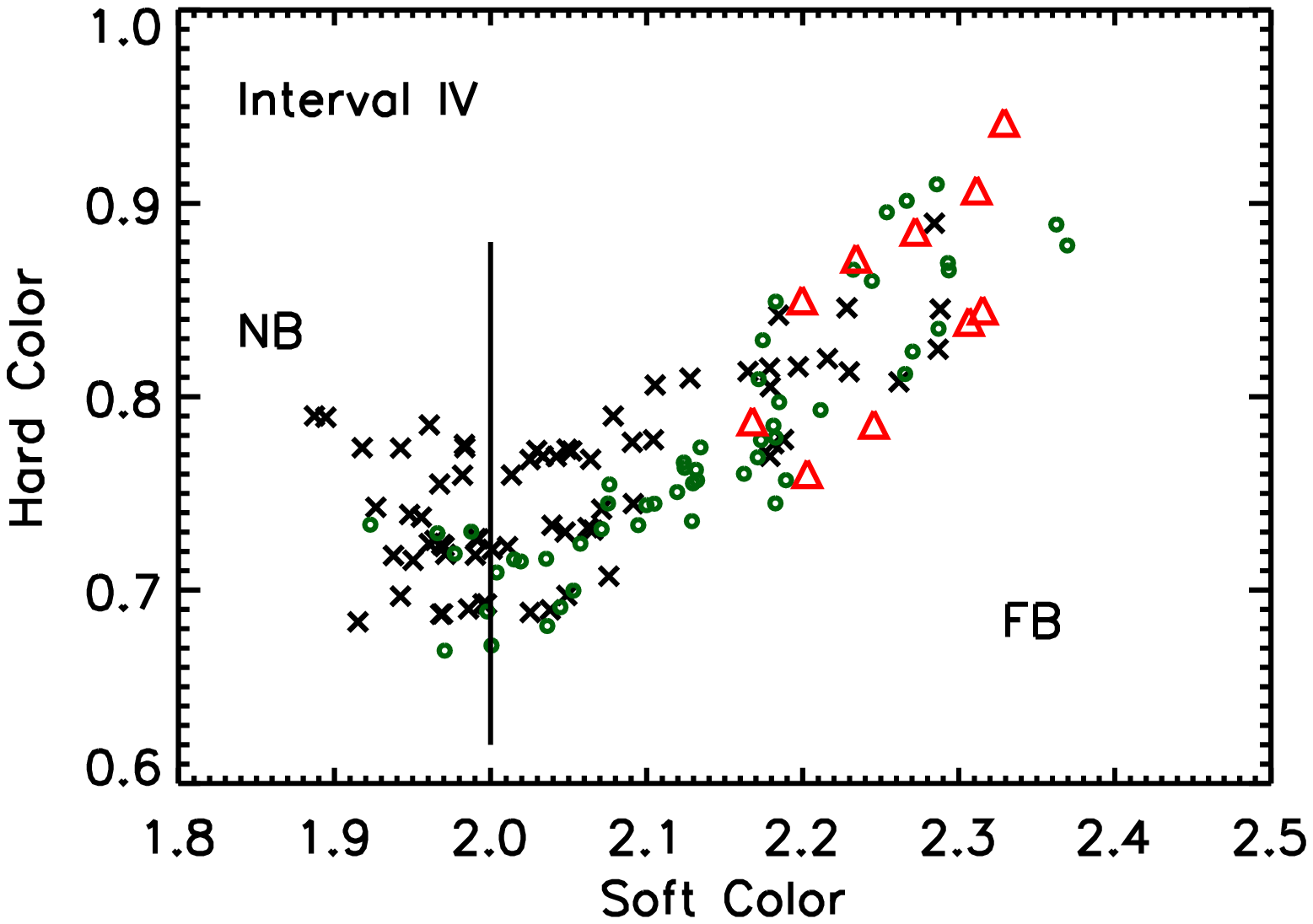}
\vspace{-2.0em}
\includegraphics[width=7cm]{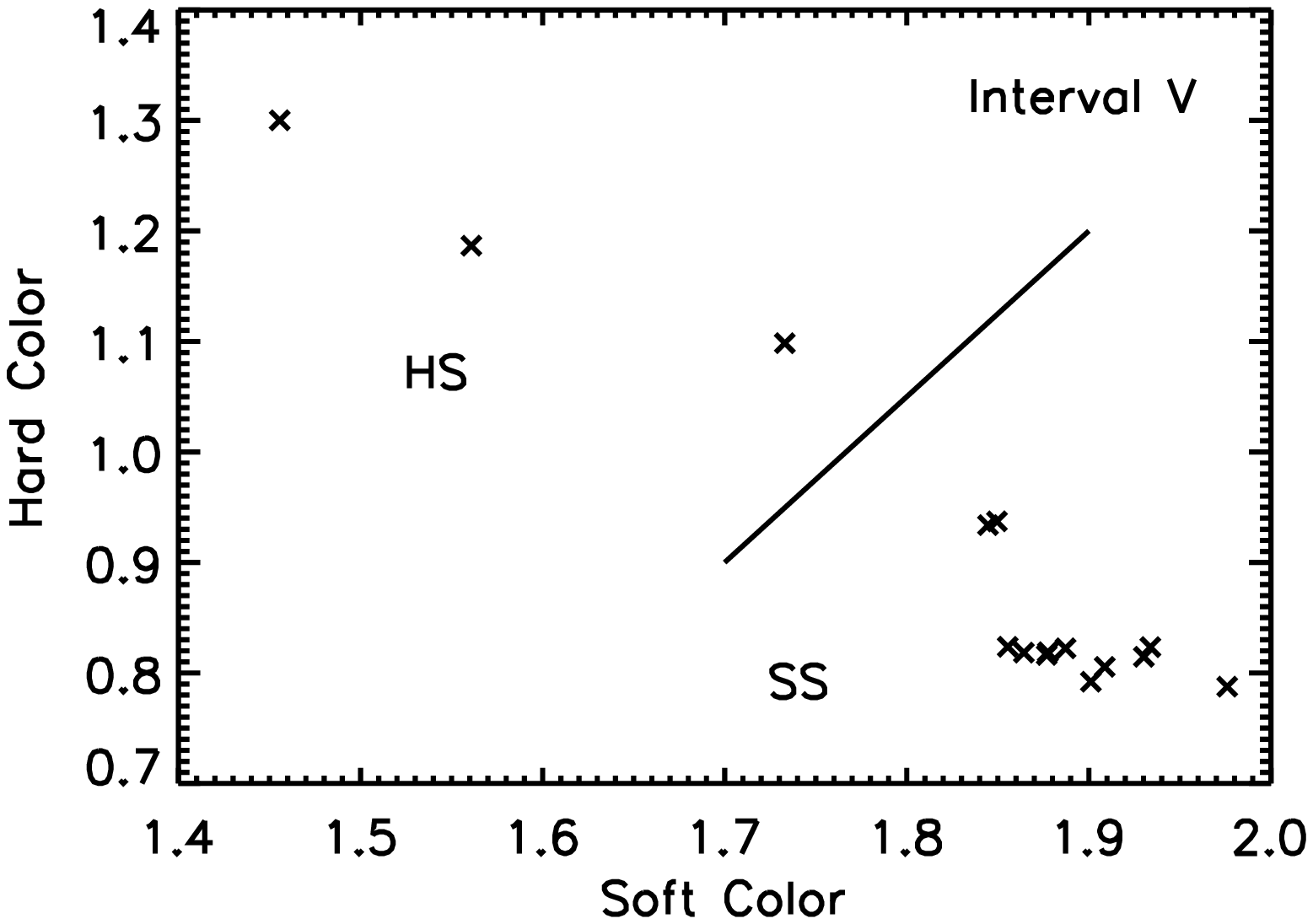}

\caption{The CCDs of the five intervals of XTE~J1701-462.
The red triangles, black crosses and green asterisks represent the data with anti-correlations, ambiguous correlations, and positive correlations, respectively. These solid lines are applied to
divide the tracks of the CCDs.}
\label{fig2}
%\end{center}
\end{figure}

\section{RESULTS}
\subsection{The distribution of cross-correlation coefficient on CCDs}

From our systematical analyses for all the observations during the 2006-2007 outburst of XTE~J1701-462, the anti-correlations and positive correlations between the soft and hard light curves are found in 51 and 120 observations, respectively, and the ambiguous correlations or weak correlations are detected in some other observations. The parameters of the observations with anti-correlations and positive correlations are listed in Table 1 and Table 2, respectively.

\begin{table*}
\caption{Log of all observations ($>$ 2000 s) which show anti-correlations.}
\begin{center}
\renewcommand{\arraystretch}{1.0}
\begin{tabular}{l c c c c c c c c}
\hline \hline
      ObsID& Date& Location& CCC& Time Lag (s)& Hardness Ratio & Pivoting Energy (keV)  \\
\hline
& & & Interval I & & & \\

\hline
91106-01-07-00&2006-01-22&HB \& upper NB&-0.14$\pm$0.01& 380.74$\pm$15.82& 1.77/0.58 &6.74$\pm$0.05\\
91106-01-12-00-a&2006-01-24&HB \& upper NB&-0.21$\pm$0.01&187.16$\pm$13.12&1.80/0.59 &5.69$\pm$0.05\\
91106-02-03-14&2006-02-06&upper NB&-0.20$\pm$0.01& -314.89$\pm$21.93&  1.74/0.56 &4.66$\pm$0.05\\

\hline

& & & Interval II & & & \\

\hline

91442-01-03-10&2006-03-02&FB&-0.28$\pm$0.04&  73.26$\pm$13.77& 1.94/0.59 &3.09$\pm$0.05\\
92405-01-01-01&2006-03-04&upper NB&-0.26$\pm$0.01&    -96.19$\pm$26.43& 1.86/0.68&4.70$\pm$0.05\\
92405-01-01-03&2006-03-06&lower NB&-0.12$\pm$0.01&  100.93$\pm$26.55&1.68/0.54 &4.68$\pm$0.06\\
92405-01-04-07&2006-03-27&FB&-0.23$\pm$0.02&-609.96$\pm$22.96&1.83/0.58&2.91$\pm$0.05\\
92405-01-05-10&2006-04-02&upper NB&-0.25$\pm$0.02&    420.48$\pm$20.08&1.85/0.65 &3.08$\pm$0.06\\
92405-01-09-02-a&2006-04-29& FB &-0.20$\pm$0.01&    380.33$\pm$16.90& 1.92/0.59&2.85$\pm$0.07\\
92405-01-09-06&2006-05-03&FB&-0.22$\pm$0.01&   -124.55$\pm$26.19& 2.09/0.66&3.12$\pm$0.05\\
92405-01-09-09&2006-05-04&lower NB&-0.11$\pm$0.01&-267.25$\pm$29.64& 1.87/0.61 & 3.52$\pm$0.06\\
92405-01-10-03-a&2006-05-06&FB&-0.23$\pm$0.01&-1277.83$\pm$29.97&2.00/0.63&2.91$\pm$0.05\\
92405-01-10-07-c&2006-05-07&upper NB& -0.31$\pm$0.01&-200.45$\pm$14.65& 1.82/0.66&5.43$\pm$0.06\\
92405-01-10-07-d&2006-05-07&FB &-0.10$\pm$0.01&   -146.35$\pm$52.30& 1.82/0.58 &3.97$\pm$0.13\\
92405-01-11-04&2006-05-15& lower NB&-0.15$\pm$0.01&447.32$\pm$9.02&1.76/0.60&2.45$\pm$0.17  \\
92405-01-11-08&2006-05-12&FB&-0.45$\pm$0.02& -7.98$\pm$14.82&1.92/0.57&3.54$\pm$0.05\\
92405-01-13-09&2006-06-01&FB& -0.24$\pm$0.02&116.34$\pm$11.10&2.11/0.68&3.55$\pm$0.05\\
92405-01-14-01&2006-06-08&upper NB& -0.35$\pm$0.01&   -164.91$\pm$24.83&1.89/0.72&8.37$\pm$0.05\\
92405-01-15-03-c&2006-06-12&FB&  -0.26$\pm$0.01&   -103.06$\pm$19.05& 1.98/0.60&4.86$\pm$0.05\\
92405-01-17-01&2006-06-25&FB&-0.15$\pm$0.02&110.67$\pm$15.24&2.03/0.62&3.44$\pm$0.05\\
92405-01-19-00-b&2006-07-08& FB &-0.28$\pm$0.02&    -28.83$\pm$15.07&  2.09/0.68&3.72$\pm$0.05\\
\hline

& & & Interval III & & & \\

\hline

92405-01-20-00&2006-07-14&upper \& lower NB&-0.27$\pm$0.02&172.23$\pm$18.21&1.83/0.65&4.92$\pm$0.06\\
92405-01-20-17&2006-07-20&upper NB&-0.18$\pm$0.01& 16.58$\pm$26.27& 1.80/0.62&4.41$\pm$0.08\\
92405-01-20-18&2006-07-16&FB&-0.12$\pm$0.01&   -500.09$\pm$29.06& 1.92/0.59&3.26$\pm$0.05\\
92405-01-25-03&2006-08-21&FB&-0.13$\pm$0.01&   -730.50$\pm$28.69&  2.01/0.64&3.03$\pm$0.05\\
92405-01-25-07-b&2006-08-23& FB  &-0.19$\pm$0.01&-276.77$\pm$28.59&  1.84/0.60&4.82$\pm$0.07\\
92405-01-28-13&2006-09-08&HB&-0.34$\pm$0.01&    -71.05$\pm$23.31&  1.93/0.75&8.72$\pm$0.05\\
92405-01-28-10&2006-09-09&lower NB&-0.27$\pm$0.01&    200.68$\pm$14.80& 1.86/0.69&4.97$\pm$0.06\\
92405-01-28-00&2006-09-09&upper \& lower NB&-0.16$\pm$0.01&     58.89$\pm$24.77&1.85/0.67&5.16$\pm$0.05\\
92405-01-28-09&2006-09-09&upper \& lower NB&-0.11$\pm$0.02&  255.91$\pm$33.00& 1.84/0.66 &5.94$\pm$0.06\\
92405-01-30-00&2006-09-22&FB&-0.23$\pm$0.02&   -443.71$\pm$19.68&   2.02/0.65&3.39$\pm$0.06\\
92405-01-31-02G-a&2006-10-01&upper NB&-0.16$\pm$0.01&54.32$\pm$14.22&   1.82/0.64&5.84$\pm$0.14\\
92405-01-31-02G-b&2006-10-01&upper NB&-0.18$\pm$0.01&     -6.95$\pm$25.88&   1.82/0.64&5.43$\pm$0.06\\
92405-01-31-16&2006-10-02&upper \& lower NB &-0.19$\pm$0.02&-32.34$\pm$12.96& 1.81/0.66&7.54$\pm$0.07\\
92405-01-31-03&2006-10-03& FB  &-0.13$\pm$0.02&  178.59$\pm$8.40& 2.26/0.76& 3.04$\pm$0.07 \\
92405-01-31-04-a&2006-10-03&FB&-0.19$\pm$0.01&    378.43$\pm$17.95&  2.06/0.65&3.09$\pm$0.05\\
92405-01-34-01-a&2006-10-21&FB&-0.39$\pm$0.02&    307.94$\pm$20.41&   2.07/0.65&3.52$\pm$0.05\\
92405-01-34-05&2006-10-25&upper \& lower NB \& FB&-0.14$\pm$0.01& 56.30$\pm$18.27&  1.83/0.63&5.82$\pm$0.05\\
92405-01-35-00-a&2006-10-27&FB&-0.22$\pm$0.02&    316.43$\pm$12.52&1.93/0.58&3.38$\pm$0.05\\
92405-01-35-00-b&2006-10-27&FB&-0.23$\pm$0.03&    208.25$\pm$26.15&  2.18/0.67&4.19$\pm$0.05\\
92405-01-35-08&2006-10-31&lower NB&-0.14$\pm$0.01&    264.77$\pm$36.85& 1.90/0.71&4.00$\pm$0.12        \\
92405-01-36-02-b&2006-11-05&FB&-0.17$\pm$0.02&   1138.41$\pm$25.76&  2.09/0.66&2.96$\pm$0.07\\
92405-01-36-02-d&2006-11-05& FB  &-0.15$\pm$0.02&    101.17$\pm$25.55&  2.03/0.65&2.95$\pm$0.05\\
92405-01-36-09-b&2006-11-06&FB&-0.26$\pm$0.01&    -12.18$\pm$15.16&   2.17/0.72&3.63$\pm$0.05\\
92405-01-36-09-d&2006-11-06&FB&-0.18$\pm$0.02&     89.25$\pm$22.57&   2.01/0.65&3.72$\pm$0.06\\
92405-01-36-05&2006-11-07&FB&-0.24$\pm$0.02&    627.53$\pm$12.16&      2.22/0.73&3.26$\pm$0.05\\
92405-01-38-05&2006-11-18&FB &-0.12$\pm$0.02&   -377.64$\pm$17.30&2.08/0.64&3.49$\pm$0.05\\
92405-01-40-01-b&2007-01-13&lower NB&-0.18$\pm$0.02&     73.89$\pm$28.76& 1.85/0.67&4.28$\pm$0.06\\
92405-01-40-05&2007-01-17&HB&-0.15$\pm$0.01&       7.49$\pm$14.94&1.94/0.79&9.08$\pm$0.12\\
92405-01-42-13-a&2007-02-01&lower NB&-0.26$\pm$0.03& 128.39$\pm$11.47&      1.88/0.70&5.00$\pm$0.07\\
92405-01-43-01-a&2007-02-02& FB &-0.18$\pm$0.01&   -217.04$\pm$19.29& 1.87/0.59&3.28$\pm$0.05 \\
92405-01-50-08&2007-03-27&FB&-0.17$\pm$0.02&    756.35$\pm$31.25&   1.93/0.62&4.84$\pm$0.06\\
92405-01-53-11&2007-04-19&FB&-0.18$\pm$0.01&     63.32$\pm$23.72&  1.99/0.67&3.08$\pm$0.08\\
92405-01-54-01&2007-04-21&upper NB&-0.18$\pm$0.02&      0.90$\pm$30.61& 1.91/0.66&6.01$\pm$0.09\\

\hline

& & & Interval IV & & & \\
\hline

92405-01-58-11&2007-05-24&FB&-0.18$\pm$0.02&   -788.96$\pm$9.23&      2.27/0.81&2.52$\pm$0.07\\
93703-01-01-00&2007-07-13&FB&-0.16$\pm$0.01&   -170.27$\pm$23.50& 2.27/0.89&4.50$\pm$0.04\\
\hline \hline

\end{tabular}
\end{center}
\footnotesize{
\begin{quote}{\bf Notes.}
The date, location of the detected anti-correlation on CCDs, cross-correlation coefficient, time lag, hardness ratio, and spectral pivoting energy of each observation are listed. The Hardness ratio is defined as the ration of soft color to hard color. The letter `a' or `b', so on, represents the segment ($>$ 2000 s) of an observation, in which the anti-correlation is detected.
\end{quote}
}
\end{table*}

\begin{table*}
\caption{Log of all observations ($>$ 2000 s) which show positive correlations.}
\renewcommand{\arraystretch}{1.05}
\renewcommand{\tabcolsep}{1.0pc}
\begin{center}
\begin{tabular}{lcccccccc}
\hline \hline
      ObsID& Date& Location& CCC& Time Lag (s)& Hardness Ratio   \\
\hline
& &  Interval I & & & \\
\hline
91106-01-04-00 & 2006-01-21 &lower NB& 0.34$\pm$0.02 & -10.03$\pm$10.53 & 1.57/0.45  \\
91106-01-10-00 & 2006-01-24 &upper NB&0.29$\pm$0.02 & -166.24$\pm$11.36 & 1.78/0.58  \\
91106-01-13-00 & 2006-01-25 &FB&0.41$\pm$0.02 &-29.95$\pm$8.12 &       1.47/0.37  \\
91106-02-01-00 & 2006-01-25 &lower NB& 0.46$\pm$0.02& 78.02$\pm$19.70 &     1.61/0.47  \\
91106-02-02-09 & 2006-01-30 &HB&0.76$\pm$0.03 &38.23$\pm$18.76 &       1.77/0.66  \\
91106-02-02-10-a & 2006-01-31 &HB&0.61$\pm$0.03 &  -17.36$\pm$10.05 &     1.81/0.64  \\
91106-02-02-10-b & 2006-01-31 &HB&0.53$\pm$0.02 & -9.28$\pm$5.62 &       1.79/0.64  \\
91106-02-02-14 & 2006-02-02 &HB&0.31$\pm$0.02 &  6.09$\pm$15.66 &       1.81/0.60  \\
91106-02-02-15 & 2006-02-02 &HB&0.47$\pm$0.02 & -12.00$\pm$8.30 &       1.78/0.61  \\
91106-02-02-11 & 2006-02-02 &HB&0.70$\pm$0.03 & -29.31$\pm$19.37 &       1.71/0.63  \\
91106-02-03-06 & 2006-02-04 &HB&0.37$\pm$0.02 & -52.86$\pm$17.60 &       1.79/0.62  \\
91106-02-03-08 & 2006-02-04 &HB&0.49$\pm$0.02 & -81.24$\pm$10.67 &       1.75/0.62  \\
91106-02-03-09 & 2006-02-04 &HB&0.61$\pm$0.04 & -40.87$\pm$13.83 &       1.78/0.61  \\
91106-02-03-15 & 2006-02-05 &lower NB \& FB&0.77$\pm$0.03 & -47.89$\pm$7.61 &       1.48/0.39  \\
91106-02-04-00 & 2006-02-10 &upper NB&0.38$\pm$0.02 &5.53$\pm$12.10 &       1.68/0.51  \\
91442-01-01-00 & 2006-02-11 &HB&0.28$\pm$0.02 & 6.21$\pm$8.46 &       1.73/0.67  \\
91442-01-01-01 & 2006-02-11 &HB&0.37$\pm$0.02 & 21.25$\pm$10.72 &       1.73/0.67  \\
91442-01-01-02 & 2006-02-12 &HB&0.43$\pm$0.02 & -8.49$\pm$10.33 &       1.81/0.65  \\

\hline

& &  Interval II & & & \\
\hline
91442-01-03-00&2006-02-24&  HB  &0.15$\pm$0.01&    592.90$\pm$28.88&      1.86/0.79\\
91442-01-03-02 & 2006-02-25 &upper NB&0.18$\pm$0.01 &  133.39$\pm$24.97 &       1.91/0.73  \\
92405-01-04-03 & 2006-03-25 &lower NB \& FB&0.10$\pm$0.01 & -53.90$\pm$14.54 &   1.81/0.59  \\
92405-01-04-09 & 2006-03-27 &FB&0.27$\pm$0.02 &5.46$\pm$9.36 &       1.87/0.58  \\
92405-01-07-00&2006-04-14& FB     &0.10$\pm$0.02&    100.89$\pm$29.82&      1.94/0.59\\
92405-01-09-02-b&2006-04-29& FB  &0.24$\pm$0.01&    -11.29$\pm$13.19&      1.88/0.58\\
92405-01-09-03-a&2006-04-30& FB     &0.13$\pm$0.01&   -181.48$\pm$22.08&      1.87/0.60\\
92405-01-09-03-b&2006-04-30&  FB &0.14$\pm$0.01&   -359.91$\pm$47.26&      1.81/0.62\\
92405-01-09-07&2006-05-03& FB &0.16$\pm$0.01&   -145.22$\pm$14.48&      2.01/0.64\\
92405-01-10-00-c & 2006-05-05 &FB&0.33$\pm$0.02 &-23.75$\pm$9.06 &       2.26/0.74  \\
92405-01-10-02 & 2006-05-06 &lower NB \& FB&0.15$\pm$0.01 &104.97$\pm$15.02 &       1.87/0.60  \\
92405-01-10-03-b & 2006-05-06 &FB&0.32$\pm$0.01 &649.77$\pm$17.93 &       1.97/0.62  \\
92405-01-10-10 & 2006-05-08 &FB&0.26$\pm$0.02 &277.43$\pm$25.66 &  2.03/0.65  \\
92405-01-11-01 & 2006-05-13 &FB&0.20$\pm$0.01 & 20.87$\pm$20.49 &       1.81/0.54  \\
92405-01-11-13 & 2006-05-15 &FB&0.20$\pm$0.01 & 94.63$\pm$38.22 &       1.74/0.59  \\
92405-01-13-01 & 2006-05-26 &FB&0.25$\pm$0.01 &  171.52$\pm$13.64 &       1.96/0.65  \\
92405-01-13-03&2006-05-28& upper NB&0.11$\pm$0.01&     -0.69$\pm$10.48&      1.92/0.64\\
92405-01-13-04 & 2006-05-29 &FB&0.22$\pm$0.02 & 187.06$\pm$20.91 &       2.27/0.78  \\
92405-01-14-02 & 2006-06-03&upper \& lower NB \& FB&0.12$\pm$0.01&-444.52$\pm$26.31&1.88/0.62  \\
92405-01-14-03-a&2006-06-04& FB  &0.12$\pm$0.01&   -254.63$\pm$24.41&      1.90/0.58\\
92405-01-14-05 & 2006-06-06 &lower NB \& FB&0.17$\pm$0.01 & 260.16$\pm$23.73 &    1.77/0.58  \\
92405-01-14-06 & 2006-06-07 &lower NB \& FB&0.33$\pm$0.02 &-146.70$\pm$13.30 &    1.76/0.56  \\
92405-01-14-09&2006-06-06&upper \& lower NB     &0.12$\pm$0.01&   -173.02$\pm$23.49& 1.80/0.62\\
92405-01-15-01 & 2006-06-10 &FB& 0.15$\pm$0.01 & -489.57$\pm$23.54 &2.00/0.61  \\
92405-01-15-03-a&2006-06-12& FB     &0.12$\pm$0.01&   -124.77$\pm$22.23&      1.83/0.57\\
92405-01-15-03-b&2006-06-12& FB &0.27$\pm$0.01&   -258.87$\pm$23.14&      1.90/0.58\\
92405-01-19-00-c&2006-07-08& FB     &0.13$\pm$0.01&   -363.29$\pm$11.22&      2.13/0.68\\
92405-01-19-02 & 2006-07-09 &lower NB \& FB&0.17$\pm$0.01 &35.12$\pm$26.89 & 1.89/0.61  \\

\hline

& &  Interval III & & & \\

\hline

92405-01-19-05G & 2006-07-13 &FB& 0.26$\pm$0.02 &-14.91$\pm$11.28 & 1.95/0.62  \\
92405-01-20-19 & 2006-07-16 &FB&0.18$\pm$0.01 &126.42$\pm$17.50 &   1.92/0.60  \\
92405-01-20-12 & 2006-07-19 &lower NB&0.47$\pm$0.02 &116.86$\pm$17.88 &       1.79/0.61  \\
92405-01-20-20&2006-07-17&FB  &0.12$\pm$0.01&    -53.69$\pm$28.89&      1.83/0.58\\
92405-01-21-00 & 2006-07-21 &FB&0.26$\pm$0.01 &435.47$\pm$23.08 &1.94/0.60  \\
92405-01-22-00 & 2006-07-28 &FB&0.25$\pm$0.02 &-1333.94$\pm$14.75 &       2.20/0.74  \\
92405-01-22-09 & 2006-07-31 &FB&0.32$\pm$0.01 & 137.62$\pm$18.55 &       2.24/0.77  \\
92405-01-22-05 & 2006-08-01 &FB&0.26$\pm$0.01 &21.03$\pm$16.94 &       2.24/0.77  \\

\hline \hline
\end{tabular}
\end{center}
\footnotesize{
\begin{quote}{\bf Notes.}
The same as in Table 1, but without spectral pivoting energy.
\end{quote}
}
\end{table*}

\begin{table*}
\contcaption{}
\renewcommand{\arraystretch}{1.05}
\renewcommand{\tabcolsep}{1.0pc}
\centering
\begin{tabular}{lcccccccc}
\hline \hline
      ObsID& Date& Location& CC& Time Lag (s)& Hardness Ratio   \\
\hline

& &  Interval III & & & \\

\hline
92405-01-23-00&2006-08-04&FB &0.14$\pm$0.02&     17.18$\pm$9.50&      1.99/0.65\\
92405-01-23-02-b & 2006-08-05 &FB&0.17$\pm$0.02 &-56.84$\pm$11.32 &       2.15/0.72  \\
92405-01-24-09 & 2006-08-15 &FB&0.39$\pm$0.02 &-192.71$\pm$19.96 &  1.90/0.60  \\
92405-01-25-07-a & 2006-08-23 &lower NB&0.20$\pm$0.01 &65.42$\pm$17.47 &1.83/0.60  \\
92405-01-27-13&2006-09-07& FB  &0.13$\pm$0.01&    391.40$\pm$35.01&      1.91/0.59\\
92405-01-28-01 & 2006-09-08 &lower NB&0.19$\pm$0.01 &74.70$\pm$27.90 &       1.83/0.62  \\
92405-01-28-02 & 2006-09-14 &FB&0.30$\pm$0.01 &250.71$\pm$20.33 & 1.96/0.63  \\
92405-01-29-03-a & 2006-09-18 &FB&0.17$\pm$0.01 &91.88$\pm$18.67 &       1.90/0.62  \\
92405-01-29-03-b & 2006-09-18 &lower NB \& FB&0.18$\pm$0.01 &1299.68$\pm$23.31 &1.91/0.63  \\
92405-01-31-23 & 2006-10-04 &FB&0.24$\pm$0.01 & 60.09$\pm$22.76 &       2.01/0.64  \\
92405-01-31-05 & 2006-10-04 &FB&0.19$\pm$0.01 & 553.05$\pm$15.64 &       2.13/0.69  \\
92405-01-32-01 & 2006-10-06 &lower NB \& FB&0.30$\pm$0.01 &-635.24$\pm$16.65 &       1.87/0.64  \\
92405-01-32-04 & 2006-10-08 &FB&0.19$\pm$0.01&165.49$\pm$18.84 &    1.95/0.63  \\
92405-01-32-10 & 2006-10-12 &FB&0.18$\pm$0.01 &  68.57$\pm$22.80 &       2.25/0.75  \\
92405-01-32-09-a & 2006-10-12 &FB&0.17$\pm$0.01 &-22.75$\pm$20.62 &1.92/0.62  \\
92405-01-36-02-c & 2006-11-05 &FB&0.19$\pm$0.01&-324.69$\pm$19.45 &       2.02/0.65  \\
92405-01-36-02-e & 2006-11-05 &FB&0.31$\pm$0.02 &355.04$\pm$18.10 &       2.03/0.65  \\
92405-01-36-09-a&2006-11-06& FB& 0.35$\pm$0.02&    -80.17$\pm$13.23&      2.24/0.75\\
92405-01-36-09-c & 2006-11-06 &FB&0.35$\pm$0.02&-80.17$\pm$13.23 &2.21/0.73  \\
92405-01-36-04 & 2006-11-07 &FB&0.17$\pm$0.01 & -40.71$\pm$9.28 &       2.29/0.76  \\
92405-01-37-00 & 2006-11-10 &FB&0.16$\pm$0.01 &-45.14$\pm$9.31 &       2.21/0.73  \\\
92405-01-37-03-a&2006-11-12& FB &0.20$\pm$0.01&    271.11$\pm$20.60&      1.88/0.60\\
92405-01-37-09&2006-11-15& FB &0.19$\pm$0.01&   -449.31$\pm$19.58&      2.05/0.64\\
92405-01-38-02-a & 2006-11-17 &lower NB \& FB&0.18$\pm$0.02&-14.61$\pm$20.96 &   1.82/0.59  \\
92405-01-41-00 & 2007-01-19 &FB&0.36$\pm$0.02 &-349.46$\pm$16.59 &       2.13/0.69  \\
92405-01-41-05-b & 2007-01-22 &FB&0.21$\pm$0.01 & -48.96$\pm$9.80 &       2.01/0.66  \\
92405-01-41-05-c&2007-01-22& FB &0.16$\pm$0.01&    270.80$\pm$23.35&      1.96/0.65\\
92405-01-41-09 & 2007-01-23 &FB&0.20$\pm$0.02 &-31.52$\pm$10.61 & 1.96/0.64  \\
92405-01-42-00 & 2007-01-26 &FB&0.33$\pm$0.01 &-30.82$\pm$22.45 &  1.97/0.65  \\
92405-01-42-06-c&2007-01-31&lower NB&0.14$\pm$0.01&    -15.85$\pm$34.19&      1.84/0.60\\
92405-01-43-01-b&2007-02-02& FB  &0.22$\pm$0.02&     49.60$\pm$15.06&      1.87/0.60\\
92405-01-43-06-a&2007-02-05&lower NB &0.25$\pm$0.01&    -71.38$\pm$19.01&      1.80/0.61\\
92405-01-43-08 & 2007-02-06 &lower NB \& FB&0.13$\pm$0.01 &-65.69$\pm$23.82 &1.87/0.59  \\
92405-01-44-07-a&2007-02-13& FB   &0.18$\pm$0.01&   -120.42$\pm$24.05&      1.90/0.61\\
92405-01-44-07-b&2007-02-13& FB &0.14$\pm$0.02& -30.18$\pm$12.23& 2.00/0.62 \\
92405-01-45-03&2007-02-18&  FB &0.14$\pm$0.01&   -476.81$\pm$26.17&      1.99/0.67\\
92405-01-45-11 & 2007-02-22 &FB&0.15$\pm$0.01 &  34.57$\pm$22.02 &       2.11/0.70  \\
92405-01-46-02 & 2007-02-25 &FB&0.37$\pm$0.01 &-315.42$\pm$16.85 & 2.02/0.66  \\
92405-01-46-01-b&2007-02-24& FB &0.16$\pm$0.01&    -28.08$\pm$26.32&      1.97/0.63\\
92405-01-46-02&2007-02-25& FB &0.37$\pm$0.01&   -315.42$\pm$16.85&      2.05/0.68\\
92405-01-46-09 & 2007-02-26 &lower NB \& FB&0.23$\pm$0.01 & -238.01$\pm$19.68 &  1.88/0.62  \\
92405-01-48-07 & 2007-03-15 &FB&0.26$\pm$0.02 &-127.26$\pm$14.23 &       2.12/0.71  \\
92405-01-50-14 & 2007-03-25 &FB&0.22$\pm$0.01 & 20.65$\pm$8.80 &       2.21/0.74  \\
92405-01-50-03 & 2007-03-25 &FB&0.36$\pm$0.01 &-46.56$\pm$10.23 &       2.17/0.71  \\
92405-01-50-04 & 2007-03-26 &FB&0.20$\pm$0.01 &505.33$\pm$19.89 &  1.96/0.64  \\
92405-01-50-10&2007-03-27&lower NB &0.17$\pm$0.02&     92.65$\pm$18.33&      1.88/0.63\\
92405-01-52-05 & 2007-04-09 &FB&0.19$\pm$0.01 & -432.46$\pm$33.13 &       2.09/0.70  \\
92405-01-52-06 & 2007-04-10 &FB&0.16$\pm$0.01 &-1204.69$\pm$19.05 &       2.14/0.73  \\
92405-01-53-00 & 2007-04-13 &FB&0.17$\pm$0.01 &4.51$\pm$7.32 &       2.22/0.77  \\
92405-01-53-02 & 2007-04-15 &FB&0.25$\pm$0.01 &-62.39$\pm$13.04 &       2.26/0.79  \\
92405-01-53-05&2007-04-19& FB&0.14$\pm$0.01&   -703.98$\pm$27.27&      1.95/0.66\\
92405-01-54-00 & 2007-04-20 &FB&0.16$\pm$0.01 &-258.58$\pm$21.37 &       2.24/0.78  \\
92405-01-55-00 & 2007-04-27 &FB&0.20$\pm$0.01 &-299.55$\pm$24.97 &1.96/0.64  \\
92405-01-55-09&2007-05-03&FB &0.14$\pm$0.01&   -128.43$\pm$11.10&      2.06/0.69\\
92405-01-56-02&2007-05-06&  FB &0.15$\pm$0.01&   -218.19$\pm$15.32&      2.09/0.70\\

\hline \hline
\end{tabular}
\end{table*}

\begin{table*}
\contcaption{}
\renewcommand{\arraystretch}{1.05}
\renewcommand{\tabcolsep}{0.8pc}
\centering
\begin{tabular}{lcccccccc}
\hline \hline
      ObsID& Date& Location& CC& Time Lag (s)& Hardness Ratio   \\
\hline

& &  Interval IV & & & \\

\hline
92405-01-57-04 & 2007-05-12 &lower NB \& FB&0.16$\pm$0.02 &  4.30$\pm$21.57 &       1.99/0.67  \\
92405-01-58-12&2007-05-23&lower NB \& FB &0.13$\pm$0.01&    142.79$\pm$24.96&      2.02/0.70\\
92405-01-59-02 & 2007-05-26 &FB&0.18$\pm$0.01 & 603.63$\pm$23.66 &       2.16/0.74  \\
92405-01-60-00 & 2007-06-01 &FB&0.16$\pm$0.01 & 110.50$\pm$31.60 &       2.03/0.72  \\
92405-01-60-01&2007-06-02&  FB   &0.15$\pm$0.01&    -65.03$\pm$13.75&      2.01/0.72\\
92405-01-60-02 & 2007-06-03 &FB&0.25$\pm$0.02 & -53.18$\pm$9.70 &       2.19/0.78  \\
92405-01-61-04-a & 2007-06-10 &FB&0.40$\pm$0.02 &37.74$\pm$16.29 &  2.11/0.75  \\
92405-01-61-04-b&2007-06-10&  FB    &0.15$\pm$0.02&    -39.77$\pm$10.28&      2.23/0.82\\
92405-01-61-05 & 2007-06-11 &lower NB&0.19$\pm$0.01 & -242.74$\pm$29.81 &       1.96/0.73  \\
92405-01-61-00-a&2007-06-08& FB    &0.26$\pm$0.01&    126.28$\pm$25.14&      2.11/0.74\\
92405-01-61-15&2007-06-11&   FB   &0.17$\pm$0.01&     31.71$\pm$12.24&      2.28/0.83\\
92405-01-62-03 & 2007-06-18 &FB&0.20$\pm$0.02 & -5.85$\pm$8.51 &       2.30/0.85  \\
92405-01-63-04-a&2007-06-24& FB     &0.15$\pm$0.01&    -36.19$\pm$22.71&      2.10/0.74\\
92405-01-63-05&2007-06-25&  FB    &0.16$\pm$0.01&     52.62$\pm$24.71&      2.09/0.76\\
92405-01-63-08&2007-06-28&FB     &0.13$\pm$0.01& 35.80$\pm$25.43&  2.10/0.76  \\
93401-01-01-11 & 2007-07-03 &FB&0.22$\pm$0.01 &248.34$\pm$22.78 &       2.17/0.78  \\
93401-01-02-09-c & 2007-07-10 &FB&0.19$\pm$0.01&-117.38$\pm$15.31 &   2.20/0.84  \\
93703-01-02-01-b & 2007-07-21 &FB&0.24$\pm$0.01 &-9.48$\pm$20.33 &       2.21/0.87  \\
93401-01-02-08&2007-07-10& FB  &0.13$\pm$0.01&   -207.50$\pm$15.92&  2.27/0.89  \\
\hline \hline
\end{tabular}
\end{table*}

In order to show the CCC distribution on the CCDs, we use different colorful symbols to label different correlations in Figure 5. As demonstrated in Figure 5, the anti-correlations, positive correlations, and ambiguous correlations are shown in each of intervals \textrm{I-IV}, but only ambiguous correlations are detected in interval \textrm{V}. Moreover, in Figure 6 we plot histograms to show the percentage of the number of the segments with anti-correlation and positive correlation to the number of the total segments (listed in Tables 1 and 2) of each branch of the CCDs of intervals \textrm{I-IV}. As shown by Figures 5 and 6, the CCC distribution is changing. For interval \textrm{I}, the anti-correlations are located around the hard vertex as well as in the upper NB, and the positive correlations are mainly distributed in the HB and lower NB. When the source enters into interval \textrm{II}, anti-correlations are mostly in the upper NB, and positive correlations are mainly in the lower NB and FB. In the third CCD, namely during interval \textrm{III}, anti-correlations appear largely in the HB and upper NB, and positive correlations also appear in the lower NB and FB, which is similar to the CCC distribution in interval \textrm{II}. The anti-correlations and positive correlations of interval \textrm{IV} mainly lie in the upper and lower FB, respectively. In interval \textrm{V}, no obvious correlations are found. Therefore, we conclude that the correlation is evolving from the Cyg-like Z source interval to the Sco-like Z source intervals.

\begin{figure}
\begin{center}
\vspace{-1.5em}
\includegraphics[width=7.5cm]{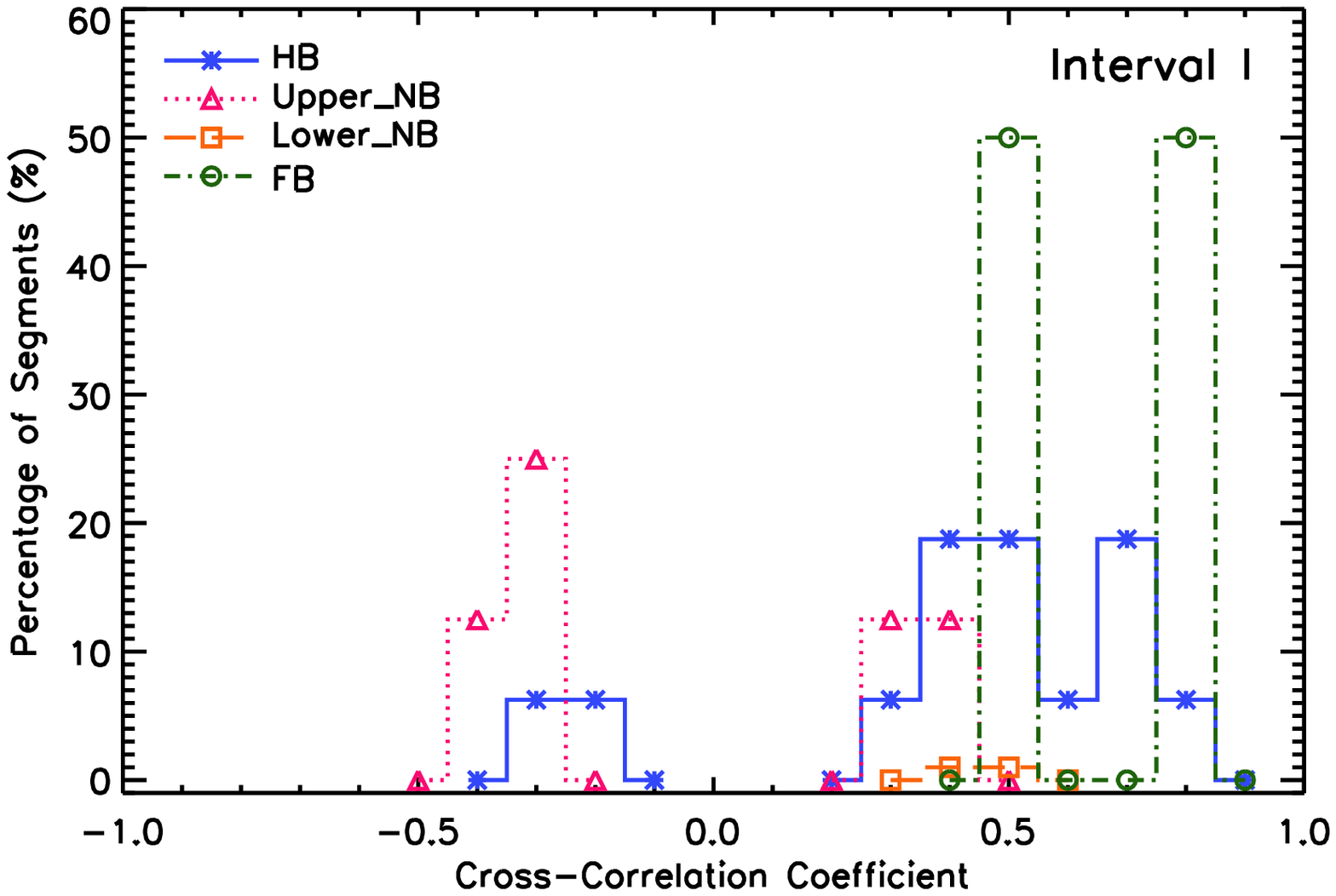}
\vspace{-1.5em}
\includegraphics[width=7.5cm]{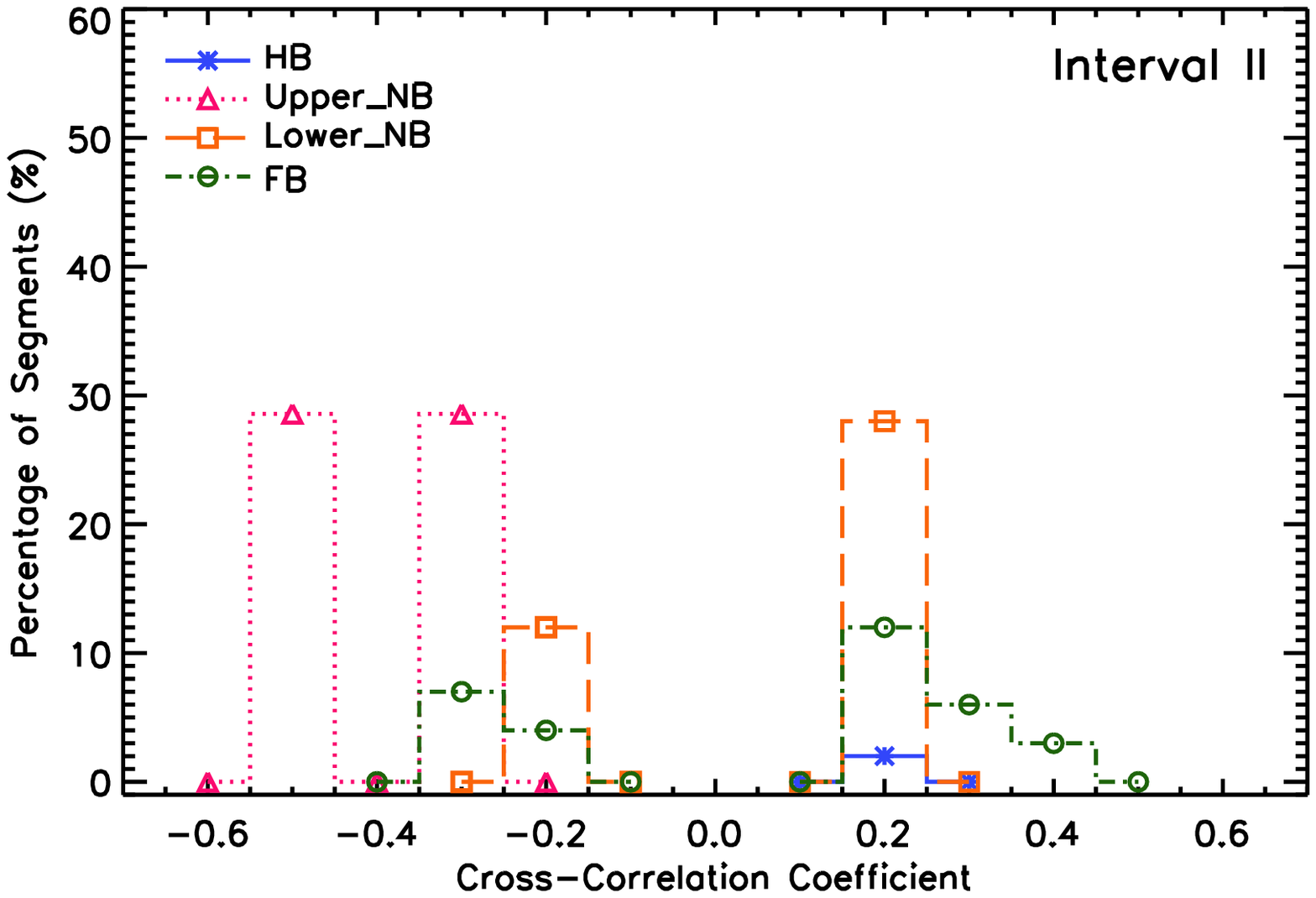}
\vspace{-1.5em}
\includegraphics[width=7.5cm]{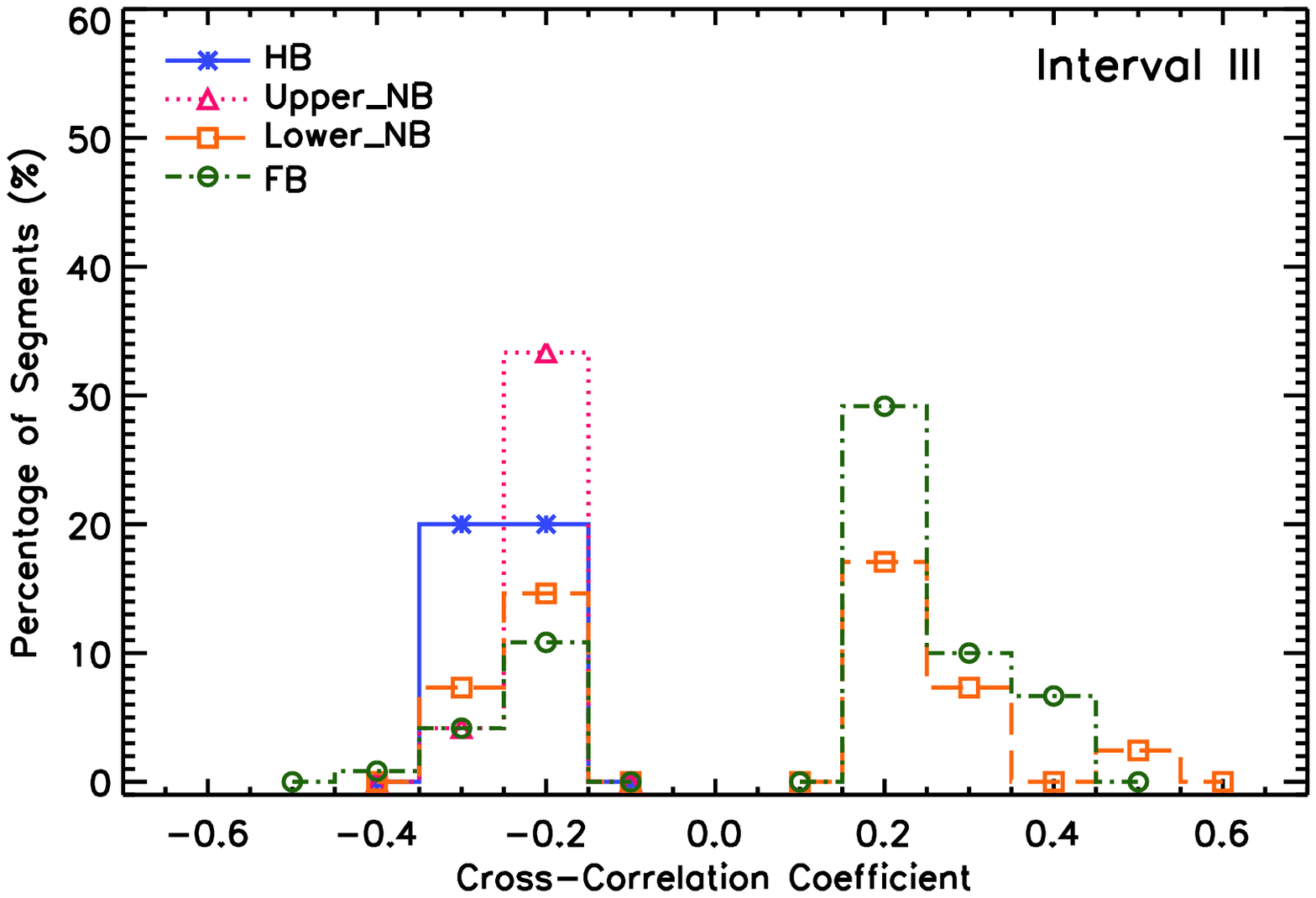}
\vspace{-1.5em}
\includegraphics[width=7.5cm]{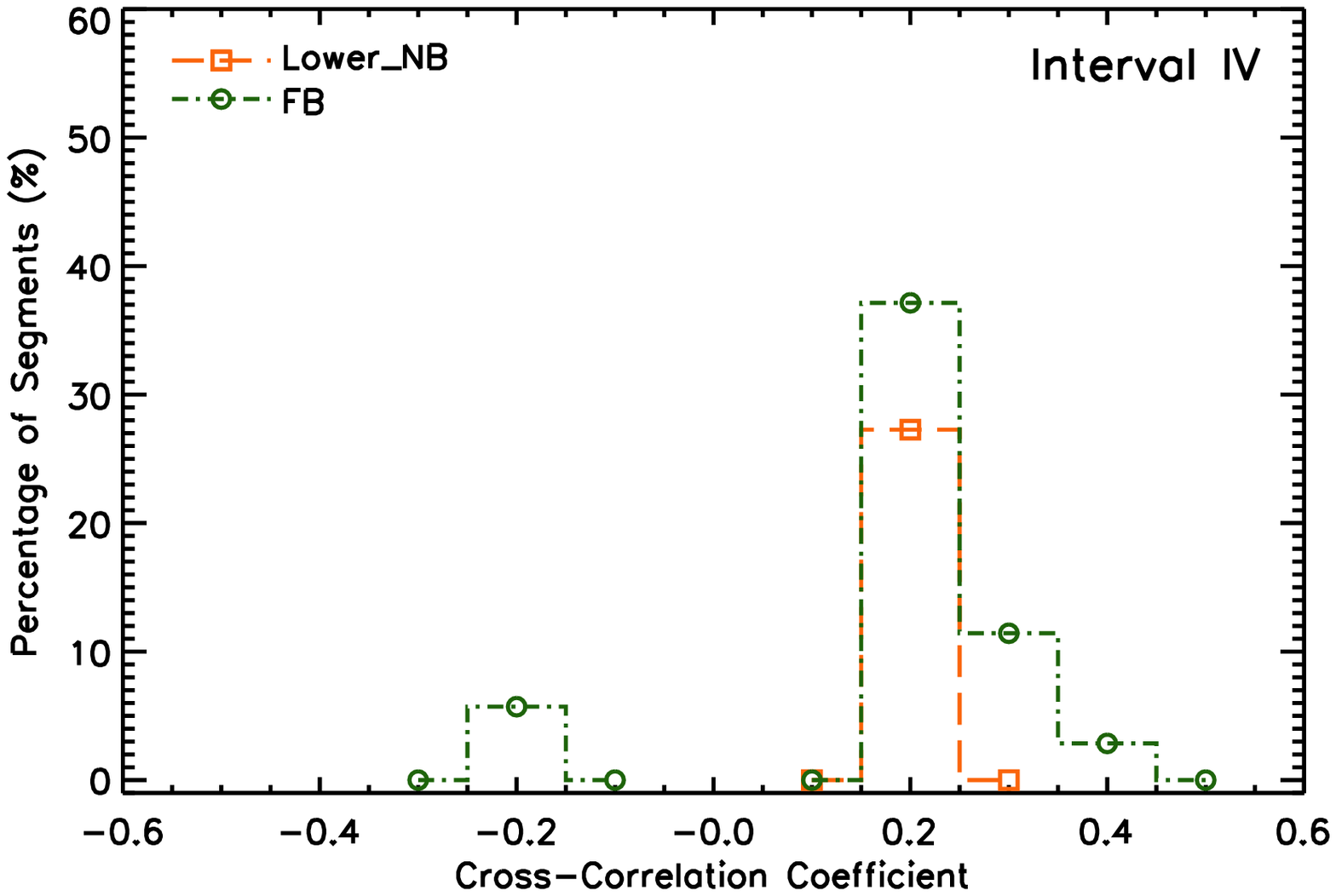}
%\vspace{-1.5em}
%\includegraphics[width=7.5cm]{paper_data/ccd_5_per.eps}
\caption{The percentage of segments as the function of CCCs.}
\end{center}
\end{figure}

 Among the 51 observations with anti-correlations, there are 27 observations with hard time lags, 22 observations with soft time lags, and 2 observations with both. As listed in Tables 1 and 2, the detected largest time lag of XTE~J1701-462 is longer than that detected in Cyg~X-2 and GX~5-1 (Lei et al. 2008; Sriram et al. 2012), but  a large proportion of lags are still lower than those reported in BHXBs (Choudhury et al. 2005; Sriram et al. 2009). The upper and low panels of Figure 7 show the segment distribution with CCC and time lag, respectively.

\begin{figure}
\begin{center}
\includegraphics[width=8cm]{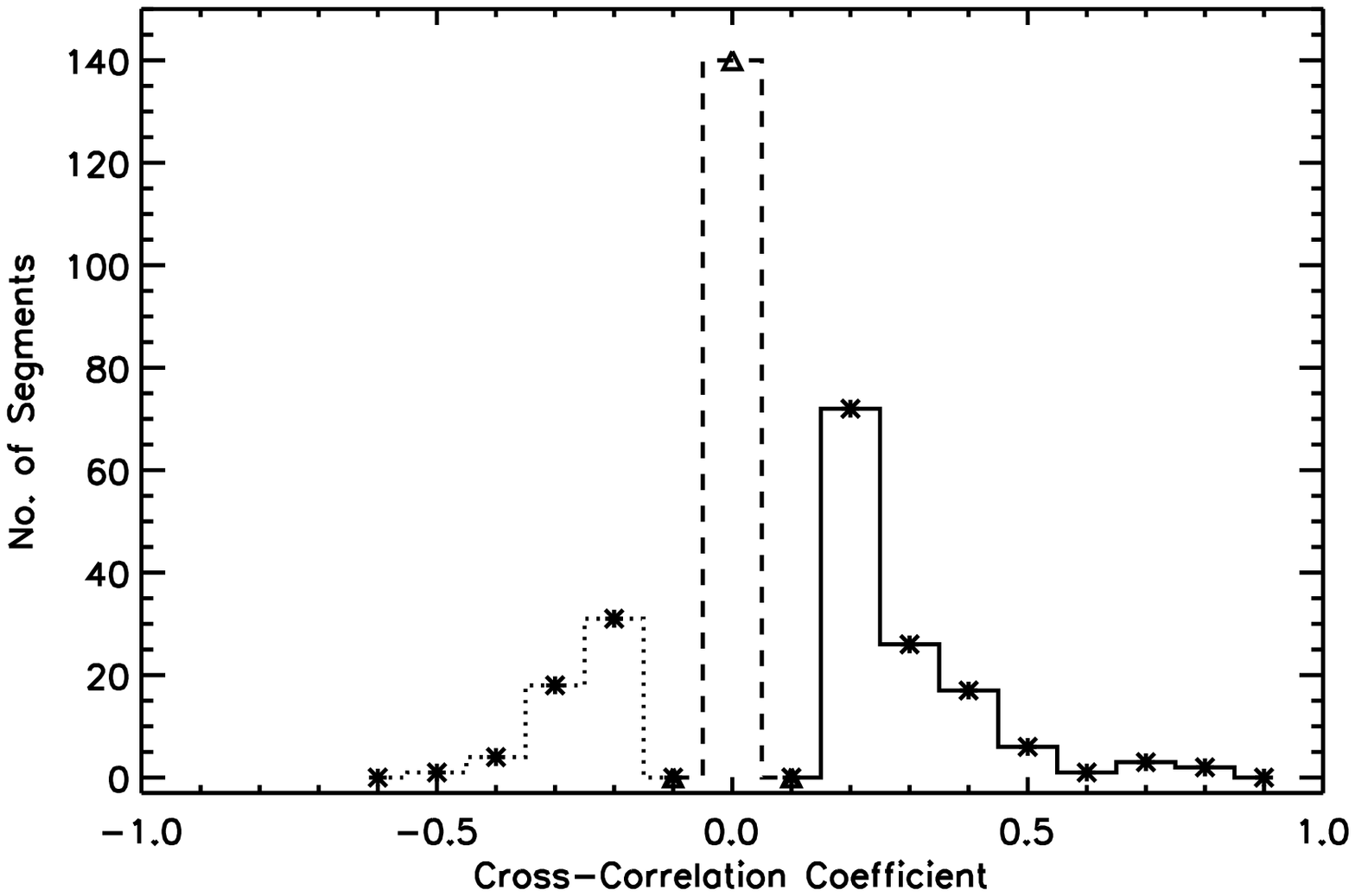}
\includegraphics[width=8cm]{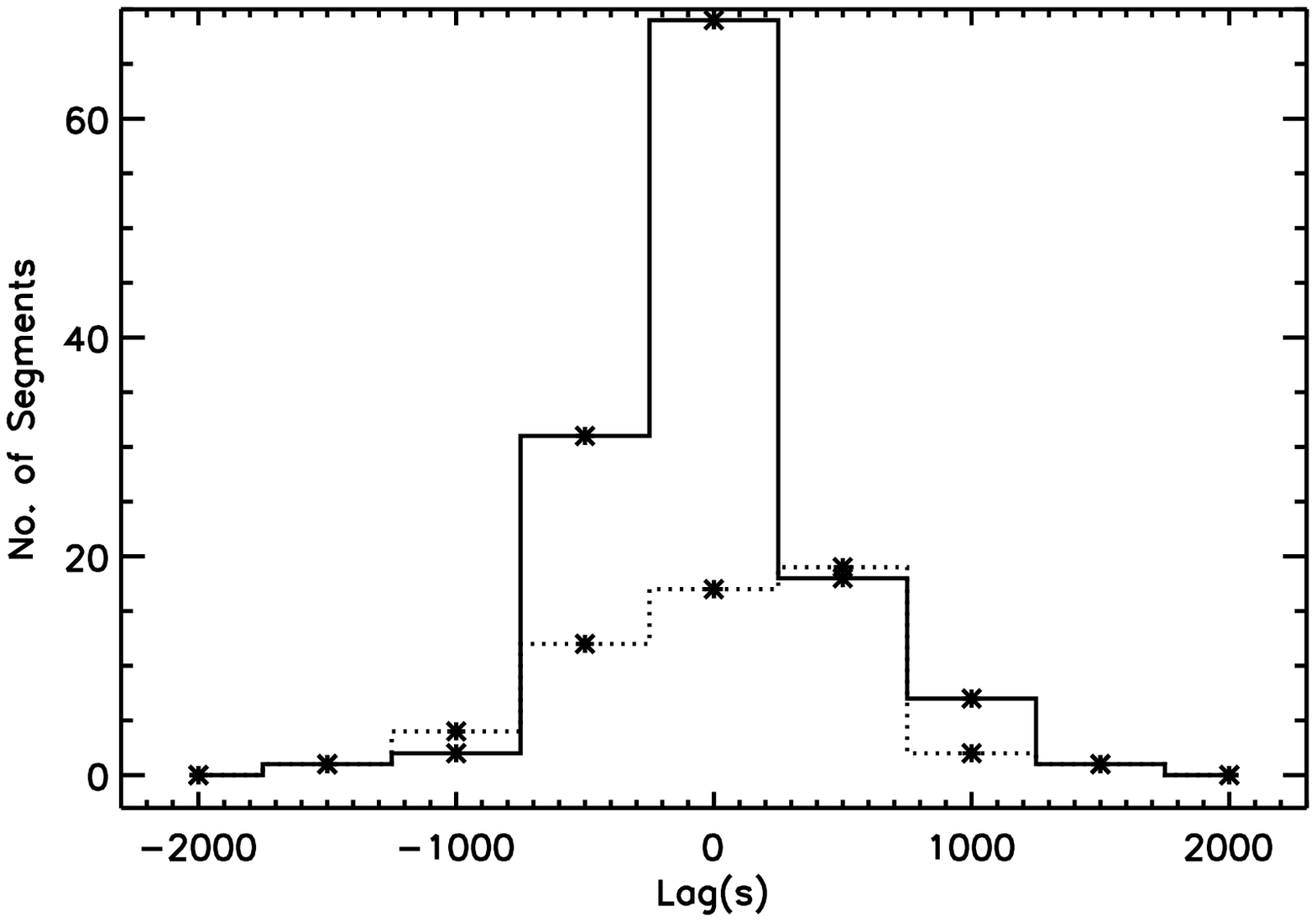}
\caption{Upper panel: the distribution of segments with CCCs, the solid line, dotted line, and dashed lines represent the CCCs from the positive correlations, anti-correlations, and ambiguous correlations, respectively. Low panel: the distribution of segments with time lags, the solid line and dotted line represent the time lags obtained from the positive correlations and anti-correlations, respectively.}
\end{center}
\end{figure}

\subsection{Spectral modeling}

   From the spectral ratios of the anti-correlated observations, we get the spectral
pivoting energies around $\sim$3-9 keV, most of which are higher in the HB and upper NB than in the lower NB and FB on the CCDs of intervals \textrm{I-III}, as listed in Table 1. Therefore, the spectral pivoting energy might be related to the position on the CCDs. The similar spectral pivoting has been found in the anti-correlated observations of several BHXBs and NS LMXBs (e.g. Choudhury et al. 2005, Sriram et al. 2007 and Lei et al. 2008). However, this kind of spectral pivoting is not found in positively correlated observations, which is similar to the result of Lei et al. (2008) and Sriram et al. (2012). This kind of spectral pivoting phenomenon might indicate that the radiative process and the spectral properties change in the anti-correlated observations.

In order to study the physical processes that drive the evolution of CCCs along the tracks on CCDs, we use different models to fit the X-ray spectra of XTE~J1701-462. Sriram et al. (2012) detected the anti-correlated time lags in the Cyg-like Z source GX~5-1 and analyzed the spectra on the HB and hard vertex of this source. In order to make accurate comparison, we chose three observations on the HB and hard vertex on the CCD in the Cyg-like interval of XTE~J1701-462 (interval \textrm{I}) to perform spectral analyses. Similarly, for comparing the spectra of XTE~J1701-462 with those of atoll source 4U~1735-44 (Lei et al. 2013), we analyze the spectra of other three observations of XTE~J1701-462 in interval \textrm{IV}, during which it approaches an atoll source. The selected observations, their positions on the CCDs, and the detected correlations of them are listed in Table 3.

We adopt the model used by Sriram et al. (2012) for GX~5-1. The model consists of a thermal Comptonization component (CompTT) developed by Titarchuk (1994), a multi-color disk blackbody (MCD, diskbb in {\it XSPEC}), and a Gaussian line component accounting for the iron emission line. The Gaussian width ($\sigma$) is fixed at 0.3 keV (Lin et al. 2009b). The fittings are statistically acceptable and the fitting parameters are listed in Table 4. The six unfolded spectra are shown in Figure 8. Additionally, in this combination model we replace the thermal Comptonization component with a single temperature blackbody (BB, bbodyrad in {\it XSPEC}), use the model of CompTT+Line+BB to fit these spectra, and get good fittings as well.

\begin{table*}
\caption{The selected observations for spectral analyses.}
\renewcommand{\arraystretch}{1.3}
\medskip
\begin{center}
\begin{tabular}{lccc}
\hline \hline
ObsID             & Interval & Position & Correlation \\
\hline
\multirow{3}*{}
91106-02-02-09    &  I       &    HB                &  positive position    \\
91106-02-02-10-a  &  I       &    HB                &  positive position    \\
91106-01-12-00-a  &  I       &    hard  vertex      &  anti-correlation     \\
\hline
92405-01-61-05    &  IV      &    soft vertex       &  positive position     \\
92405-01-57-04    &  IV      &    The botom  of FB  &  positive position     \\
92405-01-58-11    &  IV      &    The top of FB     &  anti-correlation      \\
\hline \hline
\end{tabular}
\end{center}
\end{table*}

\begin{table*}
\caption{The spectral fitting parameters of the selected observations.}
%\scriptsize{}
%\label{table:4}
%\renewcommand{\tabcolsep}{0.15pc} % enlarge column spacing
\renewcommand{\arraystretch}{1.3} % enlarge line spacing
\medskip
\begin{center}
\begin{tabular}{l c c c c c c c c}
\hline \hline
~ObsID &Interval & $kT_{in}$(keV) &  $N_{disk}$ & $kT_{e}$(keV) & $\tau$ & $Flux_{diskbb}/Flux_{total}$ & $Flux_{CompTT}/Flux_{total}$ & $\chi^2$/dof \\

\hline
\multirow{3}*{}
91106-02-02-09&I & $1.13_{-0.11}^{+0.15}$ & $379.58_{-89.71}^{+68.04}$ & $2.66_{-0.05}^{+0.04}$ & $5.67_{-0.44}^{+0.88}$ & $4.17/14.43(28.9\%)$ & $10.26/14.43(71.1\%)$ & 28/42 \\
~91106-02-02-10-a& I& $1.10_{-0.09}^{+0.10}$ & $476.68_{-58.14}^{+38.27}$ & $2.69_{-0.04}^{+0.04}$ & $5.07_{-0.24}^{+0.36}$ & $4.55/19.69(23.1\%)$ & $15.14/19.69(76.9\%)$ & 33/42 \\
~91106-01-12-00-a& I& $1.18_{-0.07}^{+0.08}$ & $483.72_{-52.10}^{+32.99}$ & $2.61_{-0.03}^{+0.03}$ & $5.02_{-0.21}^{+0.31}$ & $6.87/22.88(30.0\%)$ & $16.01/22.88(70.0\%)$ & 29/42 \\
\hline
\multirow{3}*{}

92405-01-61-05  &IV & $1.38_{-0.16}^{+0.34}$ &$55.64_{-27.62}^{+18.35}$ & $2.88_{-0.07}^{+0.08}$ & $6.04_{-0.51}^{+3.75}$ & $1.82/4.70(38.7\%)$ &
$2.88/4.70(61.3\%)$ & 37/42 \\
~92405-01-57-04  &IV & $1.23_{-0.09}^{+0.11}$ & $99.65_{-14.00}^{+11.54}$ & $2.82_{-0.07}^{+0.08}$ & $5.08_{-0.24}^{+0.28}$ & $1.72/6.62(29.6\%)$ & $4.90/6.62(70.4\%)$ & 48/42 \\
~92405-01-58-11   & IV& $1.36_{-0.10}^{+0.12}$ & $62.35_{-10.68}^{+10.06}$ &
$2.90_{-0.16}^{+0.17}$ & $6.17_{-0.23}^{+0.24}$& $1.85/7.93(23.3\%)$ &
$6.08/7.93(76.7\%)$ & 40/42 \\
\hline \hline
\end{tabular}
\end{center}
\newenvironment{mymathfrac}[2]{\raise1ex\hbox{$#1$} \left/ \lower1ex\hbox{$#2$} \right.}
% \footnote {
\footnote{ %\begin{flushleft}
\begin{quote}{\bf Notes.}
%\textit{
The model of diskBB+CompTT+Line is used and the errors of parameters are quoted at a 90\% confidence level. The flux is in units of 10$^{-9}$ ergs cm$^{-2}$ s$^{-1}$, in 3.0-25.0 keV; $kT_{in}$: inner disk temperature;
$N_{disk}$: normalization of diskBB,
$kT_{e}$: electron temperature of CompTT; $\tau$: optical depth of the Compton cloud;
$Flux_{diskbb}/Flux_{total}$: the ratio of the diskbb flux to the total flux;
$Flux_{CompTT}/Flux_{total}$: the ratio of the CompTT flux to to the total flux.
\end{quote}}
%\end{flushleft}}
\end{table*}

\begin{figure*}
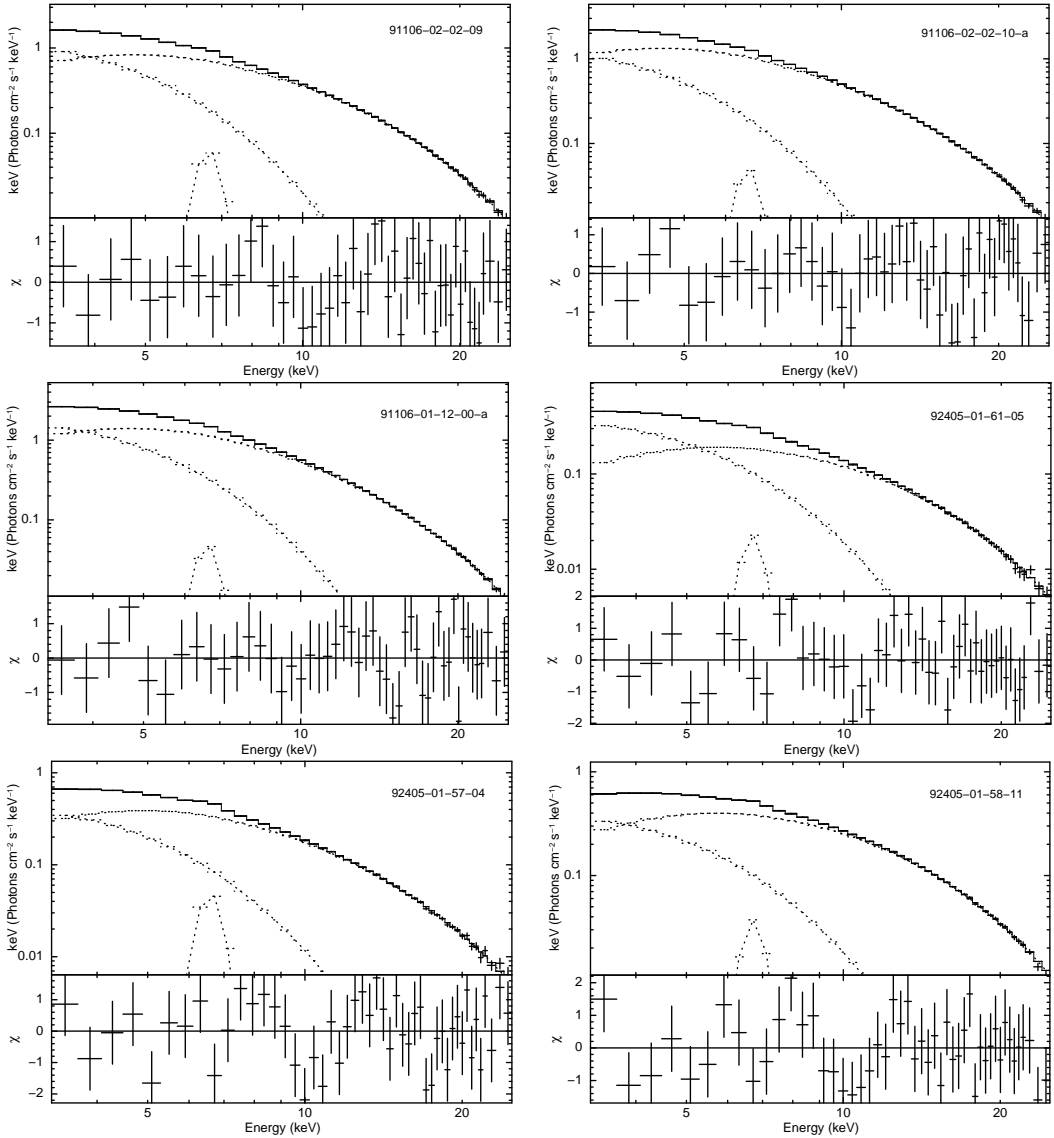

\begin{center}
\includegraphics[width=5cm,angle=270,clip]{paper_data/91106-02-02-09.eps}
\includegraphics[width=5cm,angle=270,clip]{paper_data/91106-02-02-10-a.eps}
\includegraphics[width=5cm,angle=270,clip]{paper_data/91106-01-12-00-a.eps}
\includegraphics[width=5cm,angle=270,clip]{paper_data/92405-01-61-05.eps}
\includegraphics[width=5cm,angle=270,clip]{paper_data/92405-01-57-04.eps}
\includegraphics[width=5cm,angle=270,clip]{paper_data/92405-01-58-11.eps}
\caption{The unfolded spectra and the corresponding residual distributions of the six selected observations for spectral analyses, using the model of diskBB+CompTT+Line.}
\end{center}
\end{figure*}

As listed in Table 4, from the HB to the hard vertex on the CCD of interval \textrm{I}, both disk emission flux (Flux$_{disk}$) and the flux of the Comptonization component (Flux$_{CompTT}$) increase as well as the inner disk radius ($\propto$ $N_{disk}^{1/2}$), but the ratios of individual component fluxes to the total flux (i.e. Flux$_{diskbb}$/Flux$_{total}$ and Flux$_{CompTT}$/Flux$_{total}$) basically keep unvaried. In interval \textrm{IV}, from the soft vertex to the top of the FB on the CCD, the Flux$_{disk}$
shows slight change, while the Flux$_{CompTT}$ increases distinctly. The inner disk radius firstly rises and then decreases. The percentage of the MCD flux decreases, whereas the percentage of the CompTT increases.
In any one of intervals \textrm{I} and \textrm{IV}, none of the inner disk temperature ($kT_{in}$), the electron temperature($kT_{e}$), and the optical depth ($\tau$) shows significant variation. $kT_{in}$ and $kT_{e}$ are higher in interval \textrm{IV} than in interval \textrm{I}, but the inner disk radius, and any flux, i.e. the individual flux and the total flux, are much smaller in interval \textrm{IV} than in interval \textrm{I}.

\section{ DISCUSSION }
\subsection{The position of cross-correlations on CCDs and their lags}

  In this work, using the data from {\it RXTE}, we systematically investigate
the cross-correlation of the peculiar transient source
XTE~J1701-462 during its 2006-2007 outburst. The entire outburst is divided into
five intervals according to the previous work.
In interval \textrm{I}, belonging to a Cyg-like interval,
the anti-correlations are detected in the hard vertex and upper NB,
and the positive correlations are mostly detected in the HB and lower NB.
In intervals \textrm{II-III}, we firstly obtain the anti-correlations in Sco-like intervals,
which are largely located in the HB and/or upper NB, and the positive correlations
are mainly detected in the lower NB and FB.
In interval \textrm{IV}, the anti-correlations appear in the upper FB, and the positive
ones mostly locate in the lower FB.

  The cross-correlation was studied in two Cyg-like Z sources,
Cyg~X-2 and GX~5-1, and two atoll sources, 4U~1735-44 and
4U~1608-52.
The results of two Cyg-like sources Cyg~X-2 and GX~5-1 show that,
most of the anti-correlated observations appear in the HB and upper NB (Lei et al. 2008; Sriram et al. 2012).
Our results show that, for XTE~J1701-462, the distributions of the anti-correlations in the CCDs
of both the Cyg-like and Sco-like intervals are similar with Cyg~X-2 and GX~5-1.
However, in Cyg-like interval \textrm{I}, the HB is mainly occupied by the positive
correlations, which is different from that in Cyg~X-2 and GX~5-1
(more ambiguous and anti-correlations appearing in the HB).
While in interval \textrm{IV}, the position of anti-correlations
is analogous to the atoll source 4U~1735-44 (Lei et al. 2013),
both locating at the top right of the CCD.
From the above results, the distribution of the CCCs in the CCD
shows obvious evolution during the entire outburst.

Generally, it is suggested that on the Z-track of Z sources the
mass accretion rate ({\it \.{M}})  monotonically evolves in the direction HB-NB-FB (Priedhorsky et al. 1986; Hasinger et al. 1990).
Church et al. (2006) proposed that ({\it \.{M}}) increases from the soft vertex (NB/FB
vertex) to the hard vertex (HB/NB vertex). However, Lin et al.
(2009b) suggested that the mass accretion rate of XTE~J1701-462
approximates to a constant through the Z-track, indicating that the
Z-track on the CCD cannot be primarily determined by the mass
accretion rate, as argued by Homan et al. (2010).
Though the evolution of {\it \.{M}} in one CCD is debatable,
{\it \.{M}} as a whole, corresponding to the luminosity
(6.8$\sim$0.6 $L_{Edd}$, see Lin et al. 2009b), is decreasing during the entire outburst.
It is interesting to note
that as luminosity decreases, the anti-correlations do not exist in the
lower NB in interval \textrm{I}, then appears in the lower NB in interval \textrm{II},
becomes more in amount in the lower NB in interval \textrm{III}, and finally
disappear in the lower NB in interval \textrm{IV}, showing in the upper FB (see Figures 5 and 6).
So, the results suggest that distribution of the CCCs might
depend on the luminosity of the source.

Similar to the previous results, most of the anti-correlated lags detected in XTE~J1701-462 are long timescale lags ($>$ 1 s). In the frame of Comptonization model, the low-energy photons
are scattered in the hot electrons to produce the observed hard lags
and this process would produce the short timescale lags ($\leq$ 1 s)
(Hasinger 1987; Nowak et al. 1999; B\"{o}ttcher \& Liang 1999).
There are some other timescales such as the dynamical and thermal
timescales of the accretion disk (Frank et al. 2002). However,
these timescales are also shorter than the observed time lags
in XTE~J1701-462.
The viscous timescale of the whole accretion
disk can account for the long timescale lags spanning from dozens
of days to hundreds of days.
Nevertheless, the observed time lags in
XTE~J1701-462 are much smaller than this kind of timescales of
the whole accretion disk.
Therefore, the accreting flow responsible
for the detected time lags in XTE~J1701-462 should be restricted
in a smaller region such as the inner disk edge, rather
than in the whole accretion disk. In this way, the produced time
lags could be much less than a day (Choudhury \& Rao 2004).
The difference in anti-correlated time lags imply the inner disk front movement occurs at different radii, which affects the properties of the corona (Lin et al. 2009b; Sriram et al. 2012).
This kind of anti-correlated lags have also been observed in several BHXBs
(Choudhury \& Rao 2004; Choudhury et al. 2005; Sriram et al. 2007, 2009, 2010)
and NS LMXBs (Lei et al. 2008; Sriram et al. 2012; Lei et al. 2013, 2014),
and  are considered to indicate the existence of a truncated accretion disk geometry.
For XTE~J1701-462, the anti-correlated lags are detected
in the first four intervals, appearing in
the hard vertex and upper NB, the HB and/or upper NB, and the upper FB, respectively,
which suggests the accretion disk could be also truncated in these branches.
However, the viscous model is not responsible for the soft lags.
Li et al. (2007) suggested that the fluctuation
in the inner region of the disk can disturb the outer material
and the disturbance stimulates the disk movement within thousands
of seconds, which could result in the soft times lags.

\subsection{Spectral analysis}

As pointed out above, the distribution of the positive correlation
of XTE~J1701-462 on the ¡°Z¡± track of the CCD during its Cyg-like
Z interval is different from that of Cyg-like
Z source Cyg~X-2 and GX~5-1. In order to study what results in
the difference, we use the spectral model of GX~5-1 (Sriram et
al. 2012) to fit some spectra in interval I of XTE~J1701-462.
As for GX~5-1, from the HB to hard vertex, the percentage of the
Compton component is increasing, whereas the dominant component has turned from
disk component to the Compton component (see Table 3, Sriram et al. 2012).
However, for XTE~J1701-462, from the HB to the hard vertex, although the percentage of the
Compton component shows a little change, and the
Compton component always dominates the flux.
Both the inner disk temperature and the electron temperature are lower in XTE~J1701-462 than in GX~5-1.
But the percentage of the Compton component is higher in  XTE~J1701-462 (see Table 4), indicating that more seed
photons are Compton up-scattered in XTE~J1701-462.
Moreover,
the ratio of the soft disk flux to the hard Comptonization
flux does not show remarkable variation in XTE~J1701-462,
which might results in that the HB of interval I of this source is
mainly occupied by positive correlations.

As mentioned above, the region with anti-relations on the
CCD in interval IV of XTE~J1701-462 is analogous to that of
atoll source 4U~1735-44, which motivates us to analyze some
spectra of this interval.
 From the fitting results of interval \textrm{IV}, we can see that at the highest
Flux$_{total}$, the corresponding Flux$_{CompTT}$ is not the lowest,
in other words, the spectrum is not the softest (see Table 4).
Therefore, the above spectral fitting result is similar to that of 4U~1735-44,
suggesting that the anti-correlations in interval \textrm{IV}
occur at a transition state.
Based on the spectral analysis and the distribution of
CCCs of the
fourth interval, the upper FB might correspond to the atoll source of UB.
For interval \textrm{IV}, Lin et al. (2009b) group it with
intervals \textrm{I-III} as the Z intervals due to the higher
luminosity of the FB than that in other atoll sources.
However, based on the results of timing and spectral analysis,
we prefer classifying interval \textrm{IV} as an atoll source,
rather than a Z source, which is consistent with the suggestion of Homan
et al. (2010).
Therefore, the luminosity could not be the crucial factor to judge which type
a source should belong to.

\section{CONCLUSION}

Using the data from {\it RXTE} for the peculiar transient low-mass X-ray binary
XTE~J1701-462 during its 2006-2007 outburst, we systematically
study the cross-correlations between its soft and hard light
curves, and find correlations in the first four intervals,
especially the Sco-like interval studied firstly.
For XTE~J1701-462, the distributions of the anti-correlations in the CCDs of the Cyg-like and Sco-like intervals (intervals I-III)
 are similar to those of the Z sources Cyg~X-2 and GX~5-1. While that in
interval \textrm{IV}, the anti-correlation distribution on the CCD is similar to that of atoll source 4U~1735-44.
The correlation, especially the anti-correlation, is
evolving with the source luminosity.
The truncated disk model might be responsible for the detected
time lags in XTE~J1701-462.
The spectral parameters
and flux evolution in this interval IV are consistent with
those of  4U~1735-44. Therefore, we argue that the interval IV
of XTE~J1701-462 could be belonged to an atoll source phase.

\section*{ACKNOWLEDGEMENTS}

We thank the anonymous referee for her/his constructive comments which are helpful for us to improve the presentation of this paper. This work has made use of data from the High Energy Astrophysics Science Archive Research Center (HEASARC), provided by NASA/Goddard Space Flight Center (GSFC). This research is subsidized by NSFC (Grant Nos. 11143013, 11173024, 11173034, 11303047, and 11203064), 973 Program (Grant Nos. 2012CB821800, 2009CB824800 and 2014CB845800), and LCWR of CAS (Grant No. XBBS201121).

\clearpage
\end{document}